\documentclass[aps, prb, floatfix, twocolumn, notitlepage, superscriptaddress, 10pt]{revtex4-2}
\usepackage{xcolor}
\usepackage{amsmath, amsthm, amssymb, bbold}
\usepackage[normalem]{ulem}
\usepackage{cancel}
\usepackage{bm}
\usepackage{microtype}
\usepackage{hyperref}
\usepackage{setspace}
\hypersetup{colorlinks,linkcolor=blue,urlcolor=blue,citecolor=blue}
\usepackage{amsmath}    
\usepackage{amssymb}
\usepackage{graphicx}   
\usepackage{verbatim}   
\usepackage{color}      
\usepackage{hyperref}   
\usepackage[normalem]{ulem}
\usepackage{natbib}
\usepackage{fixmath}
\usepackage{enumitem}
\usepackage{dsfont}

\def \ttau {{\tau}}

\newcommand{\n}{N}
\hypersetup{colorlinks,linkcolor=blue,urlcolor=blue,citecolor=blue}
\newcommand{\beq}{\begin{equation}}
\newcommand{\eeq}{\end{equation}}
\newcommand{\bea}{\begin{eqnarray}}
\newcommand{\eea}{\end{eqnarray}}

\newcommand{\dom}[1]{{\color{black}#1}}

\newcommand{\e}{\text e}

\newcommand{\be}{\begin{equation}}
\newcommand{\ee}{\end{equation}}

\definecolor{darkgreen}{rgb}{0,0.5,0}
\definecolor{orange}{rgb}{1,0.0,0}
\definecolor{grey}{rgb}{.6,.6,.6}

\newcommand{\cpm}[1]{{\color{black}{#1}}}

\newcommand{\zg}[1]{{\color{black}{#1}}}
\newcommand{\zgn}[1]{{\color{black}{#1}}}

\newcommand{\tT}{T}

\newcommand{\bra}[1]{\langle #1|}
\newcommand{\ket}[1]{|#1\rangle}

\newcommand{\bJ}{\mathbf J}

\newcommand{\bchi}{\boldsymbol \chi}
\newcommand{\bkappa}{\boldsymbol \kappa}

\newcommand{\br}{\mathbf r}

\newcommand{\cD}{{\cal D}}

\begin{document}
\title{\zg{Collective tunneling of a Wigner necklace in carbon nanotubes}}
\author{Dominik  Szombathy}
\affiliation{Department of Theoretical Physics,  Institute of Physics, Budapest University of Technology and Economics,  Budafoki \'ut 8., H-1111 Budapest, Hungary}
\affiliation{MTA-BME Quantum Dynamics and Correlations Research Group, 
Institute of Physics, Budapest University of Technology and Economics,  Budafoki \'ut 8., H-1111 Budapest, Hungary}
\affiliation{Nokia Bell Labs, Nokia Solutions and Networks Kft, 1083 Budapest, B\'okay J\'anos u. 36-42, Hungary}
\author{Mikl\'os Antal Werner}
\affiliation{MTA-BME Quantum Dynamics and Correlations Research Group, 
Institute of Physics, Budapest University of Technology and Economics,  Budafoki \'ut 8., H-1111 Budapest, Hungary}
\affiliation{Strongly Correlated Systems 'Lend\" ulet' Research Group, 
Wigner Research Centre for Physics, P.O. Box 49, 1525 Budapest, Hungary}
\author{C\u at\u alin Pa\c scu Moca}
\affiliation{MTA-BME Quantum Dynamics and Correlations Research Group, Budapest University of Technology and Economics, M\"uegyetem rkp. 3., H-1111 Budapest, Hungary }
\affiliation{Department of Physics, University of Oradea,  410087, Oradea, Romania}
\author{\"Ors Legeza}
\affiliation{Strongly Correlated Systems 'Lend\" ulet' Research Group, 
Wigner Research Centre for Physics, P.O. Box 49, 1525 Budapest, Hungary}
\affiliation{Institute for Advanced Study,Technical University of Munich, Lichtenbergstrasse 2a, 85748 Garching, Germany}
\author{Assaf Hamo}
\affiliation{Department of Physics, Harvard University, Cambridge, MA 02138, USA}
\author{Shahal  Ilani}
\affiliation{Department of Condensed Matter Physics, Weizmann Institute of Science, Rehovot 76100, Israel.}
\author{Gergely Zar\'and}
\affiliation{MTA-BME Quantum Dynamics and Correlations Research Group, Budapest University of Technology and Economics, M\"uegyetem rkp. 3., H-1111 Budapest, Hungary }
\date{\today}
\begin{abstract}
The collective tunneling of  a Wigner necklace –  a \cpm{crystal-like} state of a small number of strongly interacting 
electrons  confined to  a suspended nanotube and subject to a double well potential –  is theoretically analyzed and compared with experiments 
in  [Shapir \emph{et al.}, Science {\bf 364}, 870 (2019)]. Density Matrix Renormalization Group 
computations, exact diagonalization, and instanton theory provide a consistent description of this very strongly interacting 
system, and show good agreement with experiments.  Experimentally  extracted and  theoretically computed tunneling amplitudes 
exhibit a scaling collapse. Collective quantum fluctuations renormalize the tunneling, and substantially enhance it  as 
the number of electrons  increases.
\end{abstract}
\maketitle

\section{Introduction}

While investigating correlation effects in electron liquids, Eugene Wigner conjectured in 1934 the existence of an electron crystal~\cite{Wigner1934}, today referred to as the \emph{Wigner crystal}. In his seminal work, 
Wigner noticed that  the interaction energy of a three-dimensional electron gas  
scales as $E_{\text{int}}\sim n^{1/3}$ with their density $n$, and dominates over the kinetic energy $E_K\sim n^{2/3}$
in the very dilute limit. 
Therefore,  electrons must become localized at very small carrier concentrations, and form a crystal. 
The kinetic energy of the electrons increases upon compression,  and the crystal  melts due to quantum 
and thermal fluctuations into an electron liquid. 
A similar solid–liquid (quantum or thermal) phase transition occurs in two spatial 
%
dimensions~\cpm{\cite{Bonsall.77}}.  In one dimension, however, quantum fluctuations  destroy long-range order, and no phase transition takes place but only a crossover between a Luttinger liquid-like state and a dilute regime with power-law crystalline correlations appears~\cpm{\cite{Schulz1993,Meyer2009}}.

Since the predictions of Wigner, tremendous effort has been devoted to detect and understand this quantum crystal. 
While  these efforts remained unsuccessful in three dimensions, Wigner crystal phases and correlations 
have been demonstrated   in two-dimensional structures~\cpm{\cite{Grimes.79,Fisher.79,Williams.82,Tsui.1982,Andrei1988,Jiang1990,Goldman1990,Buhmann1991,Santos1992,Shirahama1995,Yoon1999,Chen2006,Cavaliere_2009,Zhu2010,Tiemann2014,Deng2016,Deng2019,Ma2020,Regan2020,Smoleski2021,Zhou2021,Li2021,Villegas2021,Falson2022,Hossain2022}}, 
as well as  more recently  in one  dimension~\cite{Shapir.2019,Deshpande.2008}. 

\cpm{
\zg{Spatial confinement suppresses quantum and classical fluctuations, and stabilizes the crystalline structure.}
When only few electrons are confined within a limited spatial region, one often refers to  \emph{Wigner molecules}. 
Such \zg{'molecules'} exhibit highly correlated arrangements due to the dominance of electron-electron 
interaction~\cite{Wendler.1996,Egger.1999,Ellenberger.2006,Yannouleas_2007,Corrigan.2021}, and are 
characterized by significant \zg{variations} in electron density as well as distinctive peaks in the density-density correlation function~\cite{Maksym_2000,Jauregui_1993,Yannouleas.2022}. In \zg{two-dimensonal} quantum dots, the formation of Wigner molecules has been observed \zg{indirectly} by various measurements, including transport~\cite{Tarucha.1996,Kristinsdottir.2011} and spectroscopy~\cite{Mintairov.2019}. 
}

\dom{The real-space structure of small crystals  became directly accessible recently. 
The structure of a small one-dimensional Wigner crystal in a carbon nanotube has been carefully probed in Ref.~\cite{Shapir.2019}, and a \emph{collective tunneling} of the crystal has been observed. 
Very recently, one-dimensional  Wigner crystals have also been observed in van der Waals 
heterostructures~\cite{Li2024}, where the artificial stacking technique used introduces 
 strain variations, and leads to the formation of domain walls, which exhibit 
unique electronic behavior, associated with the formation of a phase-locked 
one-dimensional Wigner crystal.  By tuning the electron density, a transition can be observed 
from a one-dimensional Wigner crystal to a dimerized Wigner crystal, and
eventually to a weakly interacting Luttinger liquid \cite{Li2024}. In this work, we focus on this latter  phenomenon, and model and analyze the tunneling of a small, one-dimensional Wigner crystal, which we refer to as the 'Wigner necklace'.
}

\begin{figure}[t!]
    \begin{center}
    \includegraphics[width=0.9\columnwidth]{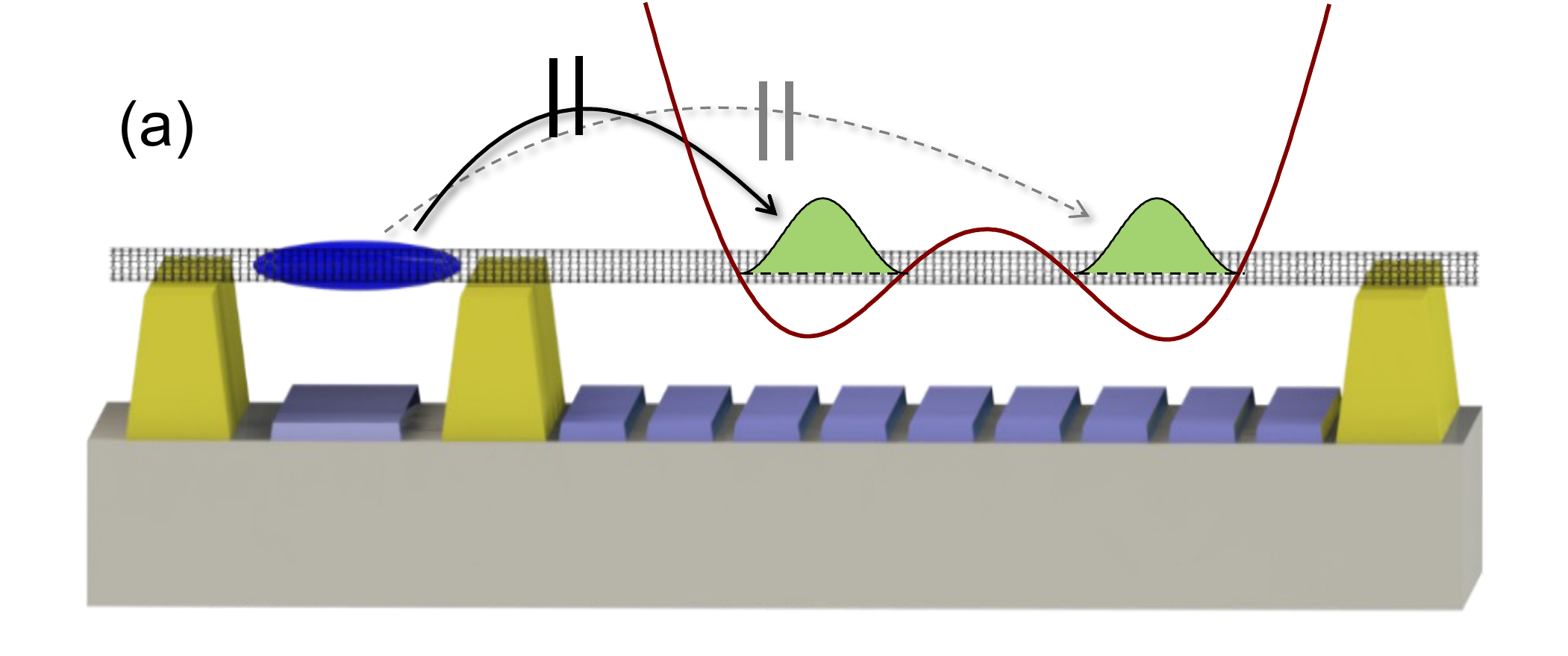} 
	 \includegraphics[width=0.49\columnwidth]{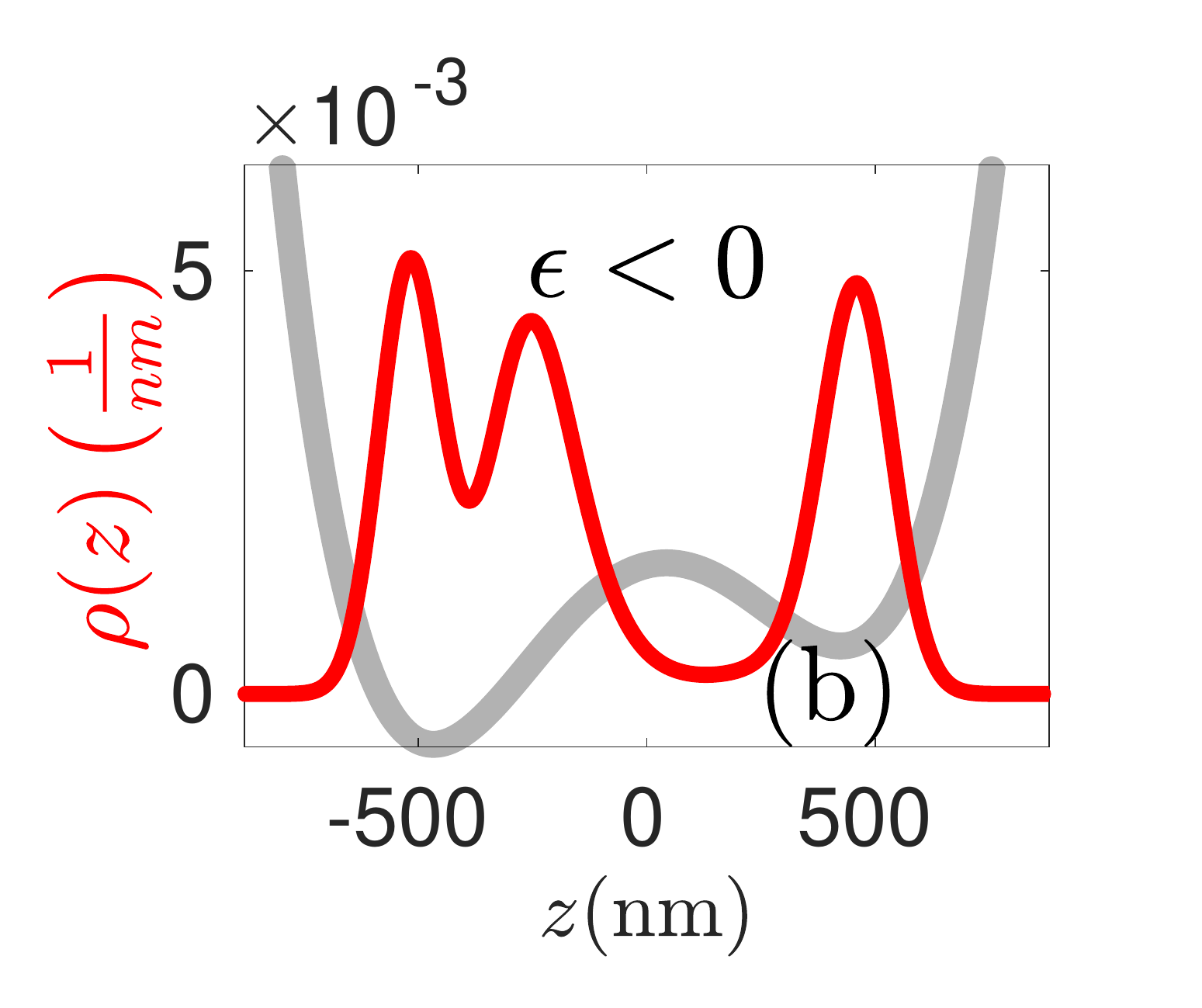}
     \includegraphics[width=0.49\columnwidth]{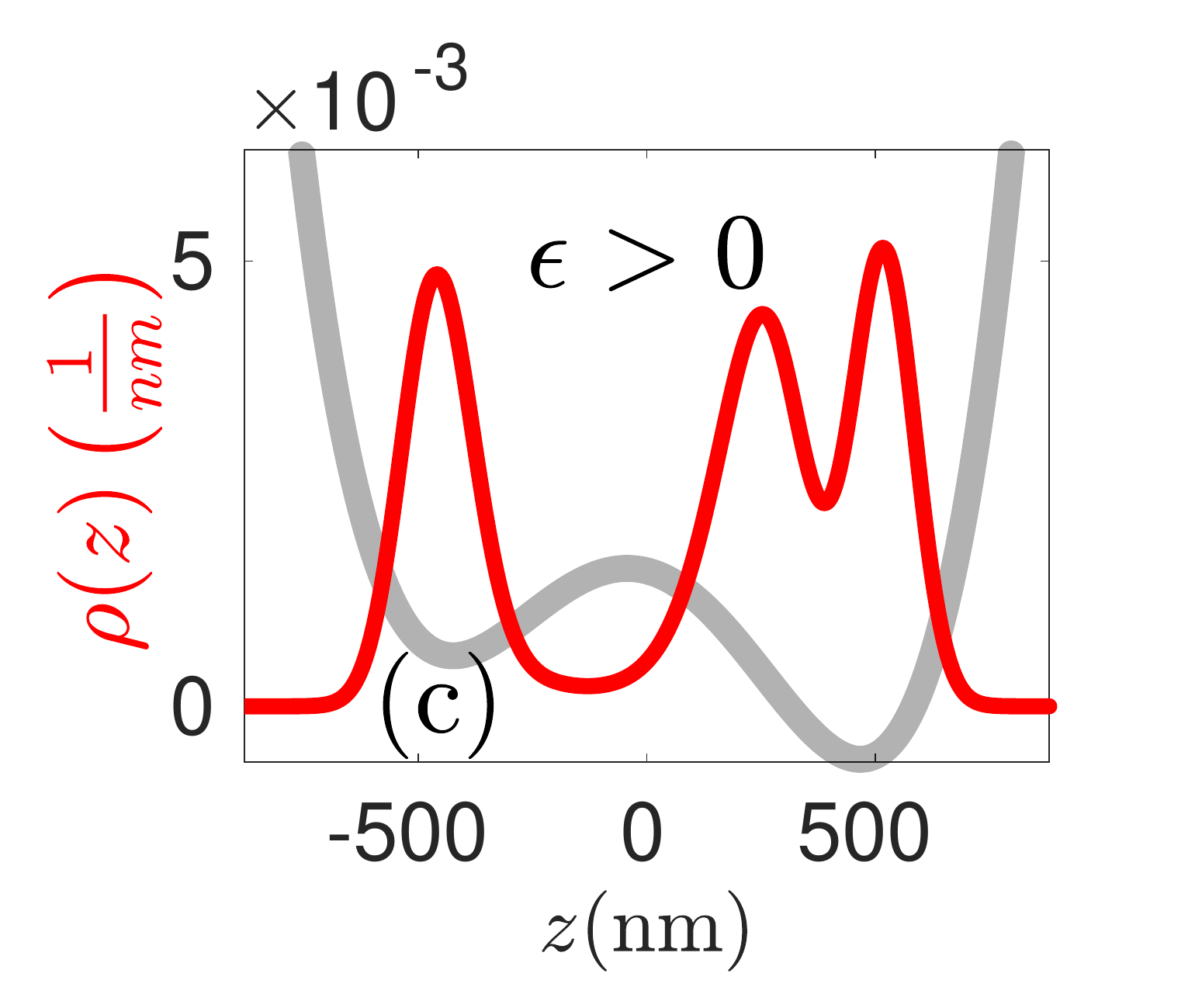}
	
    \end{center}
    \caption{(a)   Experimental setup of Ref.~\cite{Shapir.2019} used to detect 
	 collective tunneling. The bottom gates under the nanotube  are used to shape the double well potential, 
	 the left-hand-side quantum dot (dark blue)  serves as a charge detector.
	 (b) and (c) Ground state charge densities (red) from exact diagonalization for a system of three electrons  in a double well potential (grey).  A single electron moves between the two sides upon changing the asymmetry. The charge density displays a crystalline structure 
	 due to the strong Coulomb interaction ($\eta=20$ in Eq.~\eqref{eq:Hamiltonian_2}) .}
     \label{fig:experimental_setup}
\end{figure}

The setup of Ref.~\cite{Shapir.2019} is illustrated in Fig.~\ref{fig:experimental_setup}(a). A carbon nanotube is suspended and appropriately gated. Gates on the right-hand side, underneath the nanotube, are utilized to trap $N$ electrons (or holes) and create a confining potential $V(z)$ at will, with $z$ being the electron coordinate along the nanotube. On the left, a quantum dot is formed from the same nanotube and serves as a charge detector. The spatial structure of charge distributions within the nanotube can be further detected by placing a probe nanotube on top of the device and measuring the charge detector's response.
The confining  potential  $V(z)$ is well approximated by a simple quartic form,
\begin{gather}
	V(z) = {b\over 4} z^4 -{a\over 2} z^2 - c \,z,
	\label{eq:potential}
\end{gather} 
with $a$, $b$ and $c$ tunable parameters. Tunneling between the two sides of the potential 
is generated by applying a bias and thereby changing the sign of $c$.


In the present work we investigate theoretically  the collective tunneling of \zgn{very} strongly interacting charged particles 
in the potential $V(z)$. Such collective tunneling occurs for an odd number of particles. Then the classical 
ground state of the particles is twofold \zg{degenerate} in a symmetrical potential, but quantum tunneling allows for the hybridization 
of these two states, and splits their energy. However, as demonstrated experimentally \cite{Shapir.2019}, due to 
the strong Coulomb interaction, moving just one charge from one side of the barrier to the other 
is accompanied by the reordering of charges and a collective motion of all particles. 

The theoretical study of this phenomenon is rather challenging in the strongly interacting regime,
where usual quantum chemistry approaches break down \cite{RontaniPRB2010}.  \zgn{In fact, we are not aware of any precision calculation in this regime, apart from our earlier work~\cite{Shapir.2019}, where a special DMRG 
procedure has been developed to compute charge density profiles. 
 Reaching the requested accuracy and obtaining the tunneling amplitude, however, is an even more difficult task.} 
We apply a combination of three 
different methods. In the deep tunneling regime,  an \emph{instanton approach} can be used \cite{Milnikov.2001}, \zg{but incorporating}
quantum fluctuations turns out to be crucial  (as well as a technical challenge \zgn{due to numerical instabilities})  in the quantum tunneling regime. 
Unfortunately, most of the experimental data turn out to be in the intermediate region, where instanton theory is inapplicable. 
To capture the physics in this  regime, too, we perform \emph{Density Renormalization Group} (DMRG) based computations, 
\zg{which we corroborate with  \emph{restricted  'exact' diagonalization} calculations}. These three approaches provide us a 
consistent  picture, are in good agreement 
 with the experimental data, and confirm the presence of collective tunneling.
 
 \zg{Through most of this work, we neglect the electron spin. While this is not justified for intermediate interactions~\cite{Yannouleas.2022}, 
 it turns out to be an excellent approximation in the strongly interacting regime studied here. By studying the spinful 
 $N=3$ case we show that in the experimentally relevant  regime of large interactions \emph{spin-charge 
 separation} takes place, where charge degrees of freedom are responsible for the collective charge tunneling, 
 while the spin degrees of freedom remain spectators at the experimentally relevant energies.}
  \zgn{By comparing the different theoretical methods and experiment, we also conclude 
  that quantum fluctuations (phonons) of a small necklace enhance  tunneling.}

 The paper is organized as follows: In Section~\ref{sec:modeling} we outline the 
basic model used to describe the experimental setup of Ref.~\cite{Shapir.2019}. 
 Section~\ref{sec:numerics} is devoted to the  discussion of the three complementary theoretical methods used in this work.
Our results are presented in Section~\ref{sec:comparison}, along with a detailed comparison with 
 the experimental data. 
\zg{ Finally, in Section–\ref{sec:spin}, we study the case of $N=3$ spinful electrons, and demonstrate spin-charge separation. }
 Our conclusions are summarized  in Section~\ref{sec:conclusions}, while  some technical details of the 
 instanton calculation are described in  Appendix~\ref{app:prefactor}.


\section{Modelling the experimental setup}\label{sec:experiment}
\label{sec:modeling} 

To observe the Wigner crystal regime in a carbon nanotube, the mass  of the charge carriers 
needs to be as large as possible, and their interaction as strong as possible. The
Wigner crystal regime is therefore ideally observed in suspended small diameter semiconducting 
nanotubes with large gaps, as the ones used in Refs.~\cite{Shapir.2019,Deshpande.2008}. 
Electrons  confined to  such nanotubes are  very well described by the effective  Hamiltonian 
\begin{gather}\label{eq:Hamiltonian_1}
H = \sum_{i = 1}^\n \left[ -\frac{\hbar^2}{2m^*}\frac{\partial^2}{\partial z_i^2} + V(z_i) \right] + \sum_{i<j}^\n \frac{e^2}{4 \pi \varepsilon_0}\frac{1}{\left| z_i - z_j  \right|},
\end{gather}
with $m^*$ the effective mass of the electrons (holes) in the nanotube, and $V(z)$ the confining potential, Eq.~\eqref{eq:potential}.
The spin $\sigma$ and the chirality $\tau$ of the particles do not appear  in this Hamiltonian~\cite{Charlier.2007}. 
They play an important role at larger electron densities~\cite{Sarkany.2017}. 
 However, since  the tunneling experiments studied here and in Ref.~\cite{Shapir.2019} are 
performed in the spin incoherent regime~\cite{Fiete2007}, \zg{in most of our analysis} we  neglect them, and 
consider simply interacting spinless fermions. 
\cpm{\zg{As we demonstrate in Section~\ref{sec:spin}, this is well justified in the large interaction limit, relevant for 
the experiments in Ref.~\cite{Shapir.2019}, where the exchange splitting is small, 
spin degrees of freedom are incoherent, and the charge sector is responsible.}}

\zg{While they turn out to be unimportant in the very strongly interacting regime studied here, 
exchange interactions as well as spin-orbit interaction do play an important 
role at \emph{intermediate densities}, where the Wigner crystal starts to melt; there they  modify the 
structure of avoided level crossings in Wigner molecules~\cite{Yannouleas.2022},  and lead to the emergence of 
magnetic correlations and magnetic phases in larger systems~\cite{Deshpande.2008,Sarkany.2017}.}

Usually, the strength of electron-electron interaction  in a homogeneous, $d$-dimensional electron gas
is characterized by the parameter $r_s=n^{-1/d}/a_B$,   the ratio of the typical distance between charge 
carriers and the Bohr radius,  $a_B = \hbar^2 \varepsilon/me^2$. At $r_s\approx 1$
the electrons' kinetic energy is approximately the same as their potential energy, while for $r_s\gg 1$ the 
interaction energy dominates. It is in the latter regime that the Wigner crystal emerges. 

In a confined potential, however, the concept of $r_s$ is not particularly useful.
There the confining potential sets a typical length scale, which in our case is simply 
the oscillator length of the quartic potential ($a = c=0$), 
\begin{equation}
	l_d = \left( \frac{\hbar^2}{m^* b} \right)^{1/6}.
\end{equation}
Introducing  the corresponding dimensionless coordinates, $z\to \chi = z/l_d$, defines then 
the natural energy scale of the problem, 
\begin{equation}
	E_0 = \frac{\hbar^2}{m^* l_d^2}\;,
\end{equation}
and leads to the definition of the dimensionless  strength of the Coulomb interaction
\footnote{As the relevant length scale is provided by the quartic part of the double well potential, 
here, we consider a  modification of the standard $r_s$ 
parameter as originally introduced by Wigner, implemented to the situation we investigate.}, 
\begin{gather}
	\eta = \frac{l_d}{a_B} = \frac{m^* e^2}{\varepsilon \, \hbar^2}\left ( \frac{\hbar^2}{m^*b}\right )^{1/6}.
\label{eq:eta}	
\end{gather}
For the nanotube investigated here and in Ref.~\cite{Shapir.2019} we obtain 
 \begin{eqnarray}
 l_d\approx 160 \,{\rm{ nm}}, \phantom{22}
 E_0 
 \approx 5.56\, {\rm{ K}}, \phantom{22}
 \eta \approx 20, 
 \end{eqnarray}
the latter signaling an extremely strong Coulomb interaction.

%


In terms of these units, we  obtain  the dimensionless Hamiltonian
\begin{gather}\label{eq:Hamiltonian_2}
	\tilde{H} = \sum_{i = 1}^\n \left[ -\frac{1}{2}\frac{\partial^2}{\partial \chi_i^2} +\frac{1}{4} \chi_i^4 - \frac{\alpha}{2}\chi_i^2 -  \epsilon \chi_i \right] + \eta \sum_{i < j}^\n \frac{1}{\left| \chi_i - \chi_j  \right|},
\end{gather}
where the dimensionless parameter $\alpha= a \,l_d^2/E_0$  sets the height of the tunneling barrier between the two valleys, while 
$\epsilon= c\, l_d/E_0$ characterizes their bias.   

\zg{In the following sections, we analyze this Hamiltonian by three complementary approaches: a semiclassical many-body tunneling approach, a DMRG-based quantum chemistry approach, and a restricted exact diagonalization method.}

%

\section{Theoretical approaches}\label{sec:numerics}

Our goal is to compute the tunneling amplitude $\Delta$ of the \zg{tiny crystal}, 
i.e., the splitting of the two almost degenerate states \zg{of the necklace} for $N=\text{odd}$, and to investigate this 
tunnel splitting and the electrons' charge distribution as a function of the  potential  height $\alpha$, and the bias $\epsilon$.
The tunneling amplitude is inversely proportional to   the polarizability of the Wigner \zg{molecule} at $T=0$ temperature, 
and is therefore directly accessible experimentally via polarization measurements, while  
charge distributions can be detected by an AFM-like method using 
a probe nanotube~\cite{Shapir.2019}.

\subsection{Instanton theory}\label{sec:IT}

\begin{figure}[b!]
    \begin{center}
    	\includegraphics[width=0.9\columnwidth]{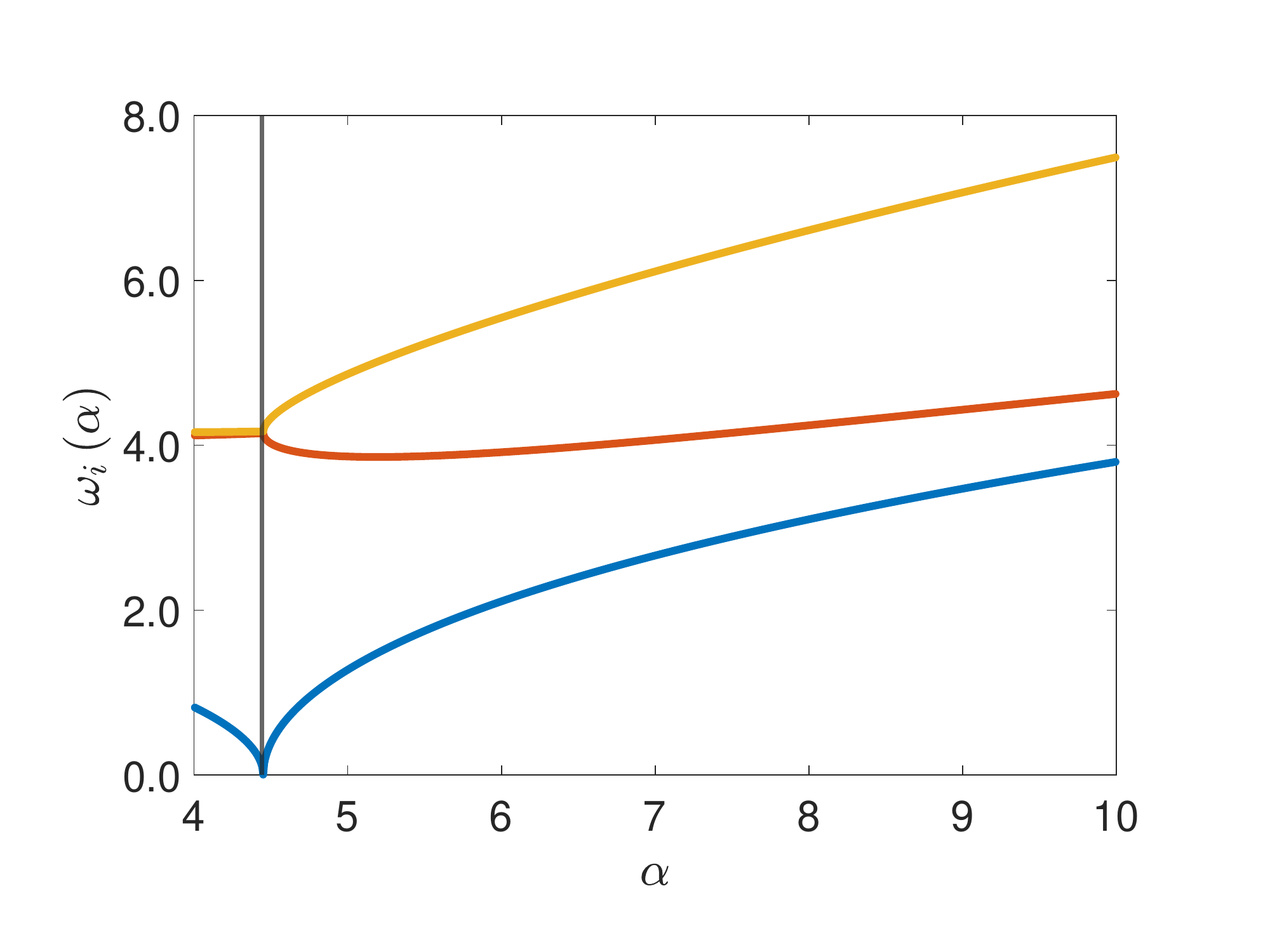}		
     \caption{Soft modes $\omega_i$ as a function of $\alpha$, for $N=3$. The vertical line at $\alpha=\alpha_{cr} \approx 4.45$ marks the beginning of the tunneling regime. For $\alpha>\alpha_{cr}$ there are two independent equilibrium position, while for $\alpha<\alpha_{cr}$ only one, indicating the absence of the tunneling at small 
	 $\alpha$. }
     \label{fig:omega}
    \end{center}
\end{figure}

We first consider the Wigner \zg{necklace}  tunneling problem by using the  instanton approach~\cite{Coleman.1988,Ciafaloni:1987qr,Grafke.2015,Richardson.2018,Milnikov.2001}.  Instanton theory  (IT)  is accurate  in 
the tunneling regime,  $\alpha\gg 0$,  however, it breaks down 
at small positive values, $\alpha\lesssim \alpha_\text{cr}$, with $\alpha_\text{cr}$ denoting the barrier height parameter, 
where tunneling sets in.

 In the instanton approach, one considers the imaginary time  tunneling amplitude between two many-body  positions. 
  Tunneling appears as a classical motion of the 
 particles in imaginary time, and the tunneling amplitude is proportional to $\Delta \sim e^{-S_\text{inst}}$, with  $S_\text{inst}$
 the instanton action. Fluctuations around this classical path determine the amplitude of tunneling, i.e.,  the prefactor 
 in front of the exponential~\cite{Coleman.1988}. 

The energy splitting $\Delta$ of the lowest lying states can be obtained by computing the imaginary time  Feynman propagator, 
\begin{gather}
K(\mathbold{\chi}_0^\prime, \mathbold{\chi}_0, \tilde{\tau}) = \langle \mathbold{\chi}_0^\prime \vert \e^{- \ttau \,\tilde{H}} \vert \mathbold{\chi}_0 \rangle,
\label{eq:Feynman}
\end{gather}
between the minima $\mathbold{\chi}_0$ and  $\mathbold{\chi}_0^\prime$
of the many-body potential, 
\begin{gather}
	v_\n(\bchi) = \sum_{i=1}^\n \left(-\frac{\alpha}{2} \chi_i^2 + \frac{1}{4} \chi_i^4   \right) + \eta \sum_{i < j}^\n \frac{1}{|\chi_i - \chi_j |},\label{eq:v_N}
\end{gather}
with $\ttau$     the dimensionless imaginary propagation time measured in units of $  \hbar/E_0$. 
One can  express  \eqref{eq:Feynman} as a path integral in terms of the imaginary time trajectories, $ \mathbold{\chi}( \ttau)$, 
which  are separated into a classical instanton trajectory, $\bchi_\text{cl}(\ttau) $, 
minimizing  the classical (Euclidean) action
\begin{equation}\label{eq:ImagAction}
	S_E[\bchi(\ttau)]  = S_0 \int_0^{T} {\rm{d}}   \ttau \Big\{ {1\over 2}\sum_{i = 1}^\n\Big( \frac{{\rm{d}} \chi_i}{{\rm{d}}  \ttau} \Big)^2 + v_\n(\bchi)   \Big \}.
\end{equation}
and small fluctuations around that,   $ \bchi(\ttau) = \bchi_\text{cl}(\ttau)  + \br(\ttau) $. 
The prefactor $S_0 = (l_d / E_0)^{3/2}$ emerges naturally, and denotes  the natural action 
unit in this problem, and ${T}$ denotes the tunneling time in units of $\hbar/E_0$. 
Expanding the action to second order in $ \br(\tau) $  leads to the expression
 \begin{gather}
	K(\bchi_0', \bchi_0, \tT) \approx 
	 e^{-S_{\text{inst}}} \int_{\br(0) = 0}^{\br(\tT) = 0} \cD\br \exp  \Big\{ \frac{1}{2} \int_{0}^{\tT} {\rm{d}} \ttau \, 
	 \br(\ttau)\cdot\nonumber \\
	  \left[ -\partial^2_{\ttau}
	 + \partial \circ 
	 \partial \, v_\n(\bchi_{cl}(\ttau)) \right] 
	 \mathbold{r}(\ttau)
	\Big \}, 
	 \label{eq:K_full}
\end{gather}
with $S_{\text{inst}} = S_E [\bchi_\text{cl}(\ttau)]$ the instanton action, and the  integral accounting for quantum fluctuations around it.

We  determined the  initial and final equilibrium positions $\bchi_0$ and $\bchi_0'$ 
as well as the instanton trajectories  by applying a Monte Carlo simulated annealing  procedure~\cite{VANDERBILT1984259}. 
Fig.~\ref{fig:omega} shows the frequencies of  small vibrations around the minimum (minima) of 
$v_\n(\bchi)$ for $N=3$. The symmetrical position of the three particles becomes classically unstable 
at $\alpha^{N=3}_\text{cr}\approx {4.45}$,  the classical 
threshold  for collective tunneling.  For $\alpha<\alpha_\text{cr}$  the minimum energy configuration is unique,  
while for  $\alpha>\alpha_\text{cr}$ two equilibrium positions exist,  and tunneling becomes possible.  
The transition to the tunneling regime is marked by the softening of the lowest energy  mode. 
Interestingly,  the direction of this mode coincides with that of the instanton trajectory
for $\alpha>\alpha_\text{cr}$.

\begin{figure}[t!]
    \begin{center}
    	\includegraphics[width=0.95\columnwidth]{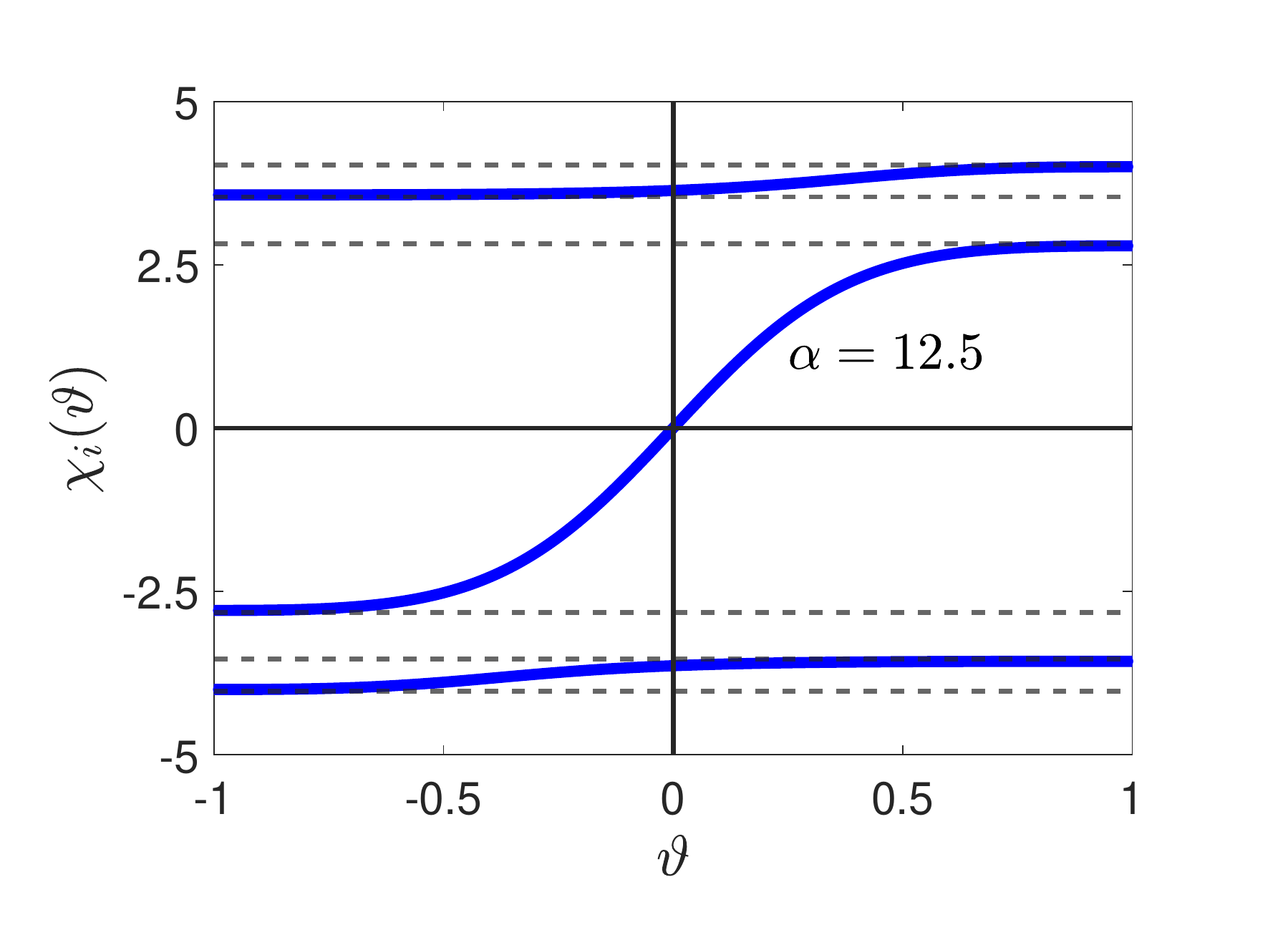}		
     \caption{ Three particle imaginary time trajectories (blue) in the dimensionless units for a specific confinement parameter $\alpha = 12.5>\alpha_\text{cr}$ and $\eta = 20$. Dashed lines show the classical equilibrium positions or instanton turning points for the 3 particles. 
     }
     \label{fig:traj}
    \end{center}
\end{figure}  

A typical trajectory is displayed for $N=3$ particles in Fig.~\ref{fig:traj}, which demonstrates 
 collective tunneling. In our calculations,  we "compactify" time by introducing the parameter 
$\vartheta = \tanh( {\ttau}/{\ttau_0})$, and parametrize $\bchi$  by using $\vartheta$.
 Clearly, the middle electron tunnels through the potential barrier, while the outer 
 electrons  do not tunnel but  adjust their positions.

The choice of $\ttau_0$ is important from the point of view of numerical accuracy. 
A small value of $\ttau_0$ increases the numerical accuracy in the tunneling region, 
while a large value of $\ttau_0\approx 5$ provides better resolution around the end of the trajectories.
Since  the primary contribution to the energy splitting arises from the region around $\ttau\approx 0$, 
a value $\ttau_0 \approx 0.5 $ turns out to be optimal for accurate calculations. 

%
%

\begin{figure}[b!]
	\begin{center}
		\includegraphics[width=0.95\columnwidth]{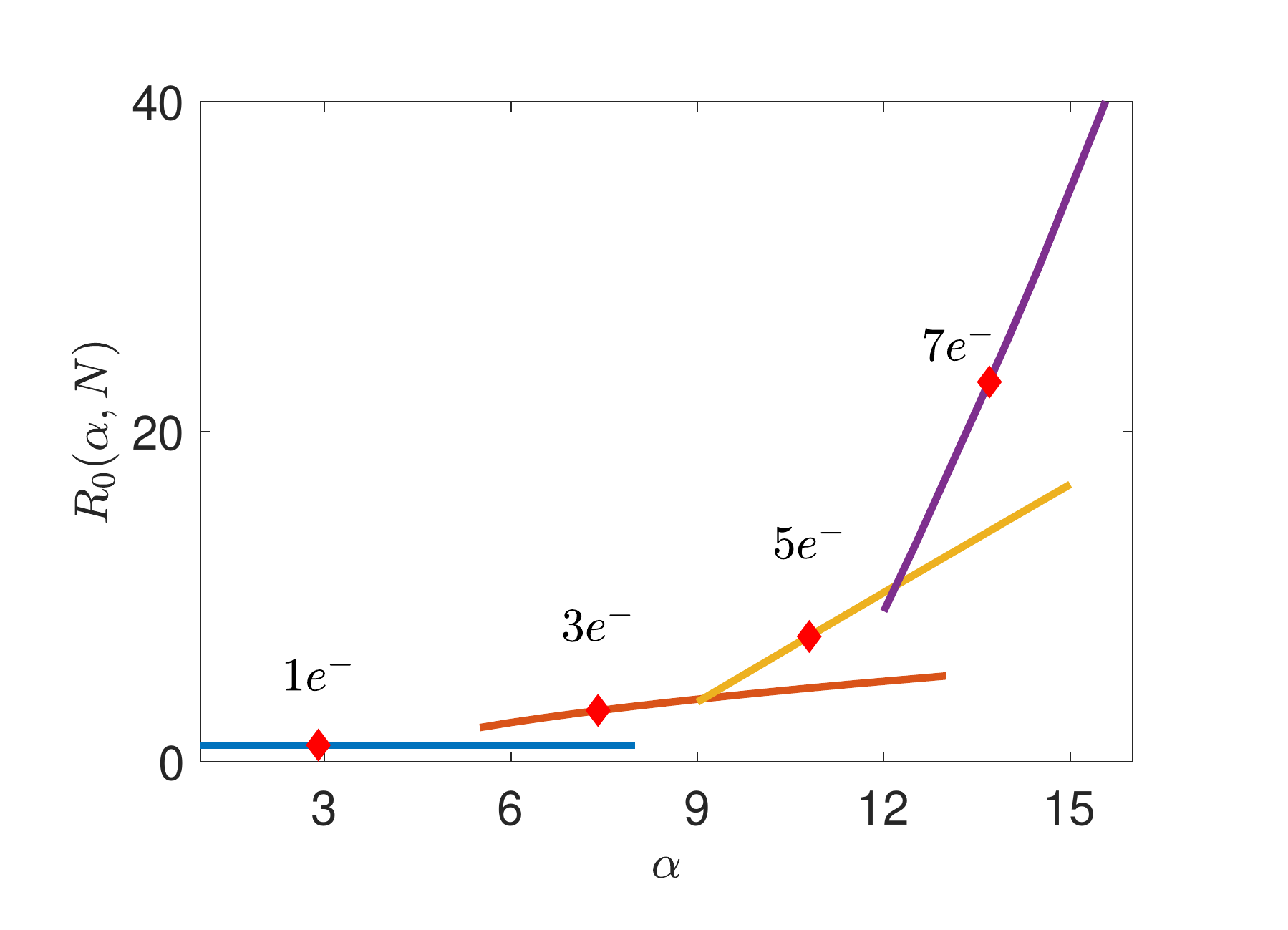}
		 \caption{
			Renormalization factor $ R_0(\alpha, N)$ as function of $\alpha$ for $N \in  \{1, 3, 5, 7\}$. 
			On each line, the diamond symbols mark the beginning of the quantum tunneling regime.  
			}
			\label{fig:R0}		
	\end{center}
\end{figure}

Performing the Gaussian integral in Eq.~\eqref{eq:K_full} is a highly non-trivial task~\cite{mackenzie2000path,Milnikov.2001}. 
The procedure consists of introducing the arc length variable ${\rm d}s^2 = {\rm d}\bchi_\text{cl}^2 $
along the instanton trajectory, and $\n-1$ perpendicular coordinates. In this way, one describes the tunneling 
as a one-dimensional tunneling process in an effective potential $w(s)$, renormalized by  'perpendicular' 
quantum fluctuations (see Apendix~\ref{app:prefactor}). 
The tunneling amplitude is then expressed as
\begin{eqnarray}
\Delta &= &R_0 (\alpha, N) \, \Delta_{1}\,, 
\nonumber 
\\
\Delta_{1}& =&  \sqrt{\frac{4\omega_{\rm{soft}}}{\pi}} \sqrt{2 [v_{\n}^{\rm{max}} - v_{\n}^{\rm{min}}]} P[{\mathbold{\chi}}_{cl}(s)] e^{-S_E},
\label{eq:instanton}
\end{eqnarray}
with $\Delta_{1}$  associated with a one-dimensional motion in the effective double-well potential, 
and $R_0 (\alpha, N)$   the aforementioned renormalization factor~\cite{mackenzie2000path,Milnikov.2001} 
(see Appendix ~\ref{app:prefactor} for details). Here $\omega_{\rm{soft}}$ denotes the oscillation frequency 
at the initial position of the tunneling trajectory in the tunneling direction, and  $P\left[ \mathbold{\chi}_\text{cl}\right]$
is a renormalization factor associated with the effective one-dimensional motion. The  prefactor $R_0 $ is  equal to 1 for $N=1$, but it   becomes significant for $N\ge 3$ (see  
 Fig.~\ref{fig:R0}), and  exhibits a non-negligible $\alpha$ dependence.  Somewhat surprisingly, 
 quantum fluctuations seem to  increase the tunneling amplitude substantially, and  
 quantitative computations  must take them  into account. 


\subsection{Density matrix renormalization group}\label{sec:DMRG}
As an alternative to instanton computations, we also performed density matrix renormalization group 
(DMRG) computations.  DMRG provides an accurate description of the intermediate tunneling 
regime, however, it fails in the deep tunneling regime, where we experience convergence problems. 

Originally, DMRG has been proposed as an efficient computational scheme 
for one-dimensional systems with short-ranged  interactions ~\cite{White.1992}, but
has  been extended later to systems with long-ranged interactions as well as to 
higher dimensional  lattices~\cite{Schollwock.2005}, and it has been 
reformulated more recently in a possibly more transparent way by using the language of 
matrix product states (MPS's)~\cite{SCHOLLWOCK201196,mcculloch2007density}. 

To perform  DMRG, we  express the Hamiltonian~\eqref{eq:Hamiltonian_2}  in a second quantized form. 
The key to efficient DMRG calculations   is to choose an appropriate basis in the strongly interacting limit studied here, 
$\eta \gg 1$.  The most natural choice of  harmonic oscillator basis functions centered at $\chi=0$, e.g., 
 is not able to reach this regime~\cite{RontaniPRB2010}. Here we perform calculations by using 
 an \emph{overcomplete adaptive basis}  with harmonic oscillator wave functions localized around the classical equilibrium 
 positions of the electrons~\cite{Shapir.2019}. 

%
%

In this basis, we  rewrite the Hamiltonian \eqref{eq:Hamiltonian_2} the second quantized form 
\begin{gather}
	\tilde H = \sum_{a,b} t_{ab} c^\dagger_a c_b + {1 \over 2} \sum_{a,b,c,d} V_{ab,cd} c^\dagger_a c^\dagger_b c_d c_c,
\end{gather}
where $t_{ab}$ stands for  the matrix elements of the non-interacting part of  \eqref{eq:Hamiltonian_2},
$t_{ab} = \langle \psi_a | \tilde H_0 | \psi_b \rangle$, while  $V_{ab,cd} = \langle \psi_a \psi_b | U | \psi_c \psi_d \rangle$ 
are  the matrix elements of the Coulomb interaction, $U=\eta/|\chi-\chi'|$ calculated within  
the single particle wave functions described above. The computation of   the matrix elements $V_{ab,cd}$ is numerically  demanding, but it can be speeded up 
by exploiting the  translational invariance of the Coulomb interaction.


\begin{figure}[t!]
    \begin{center}
    		\includegraphics[width=0.95\columnwidth]{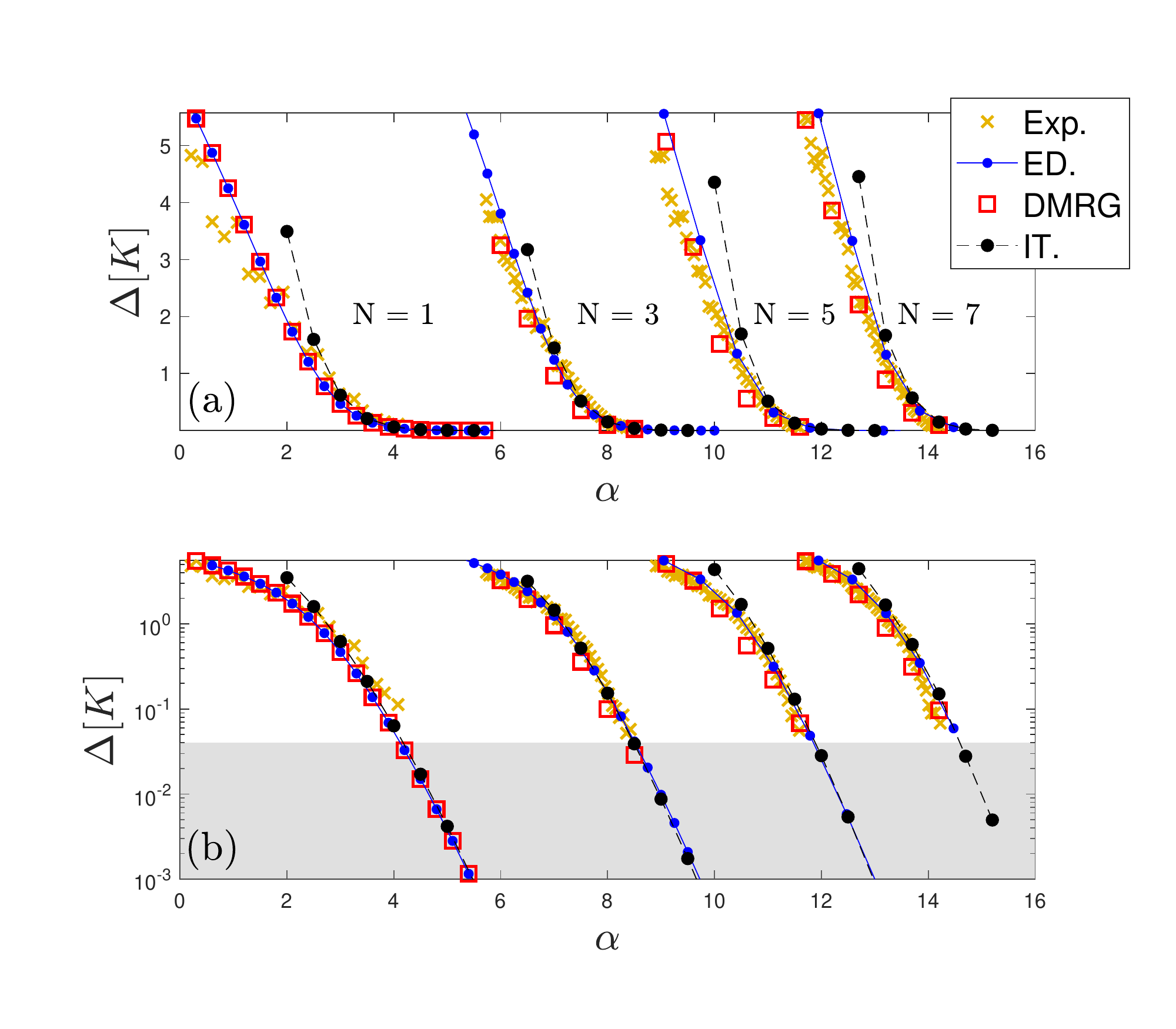}
     		\includegraphics[width=0.95\columnwidth]{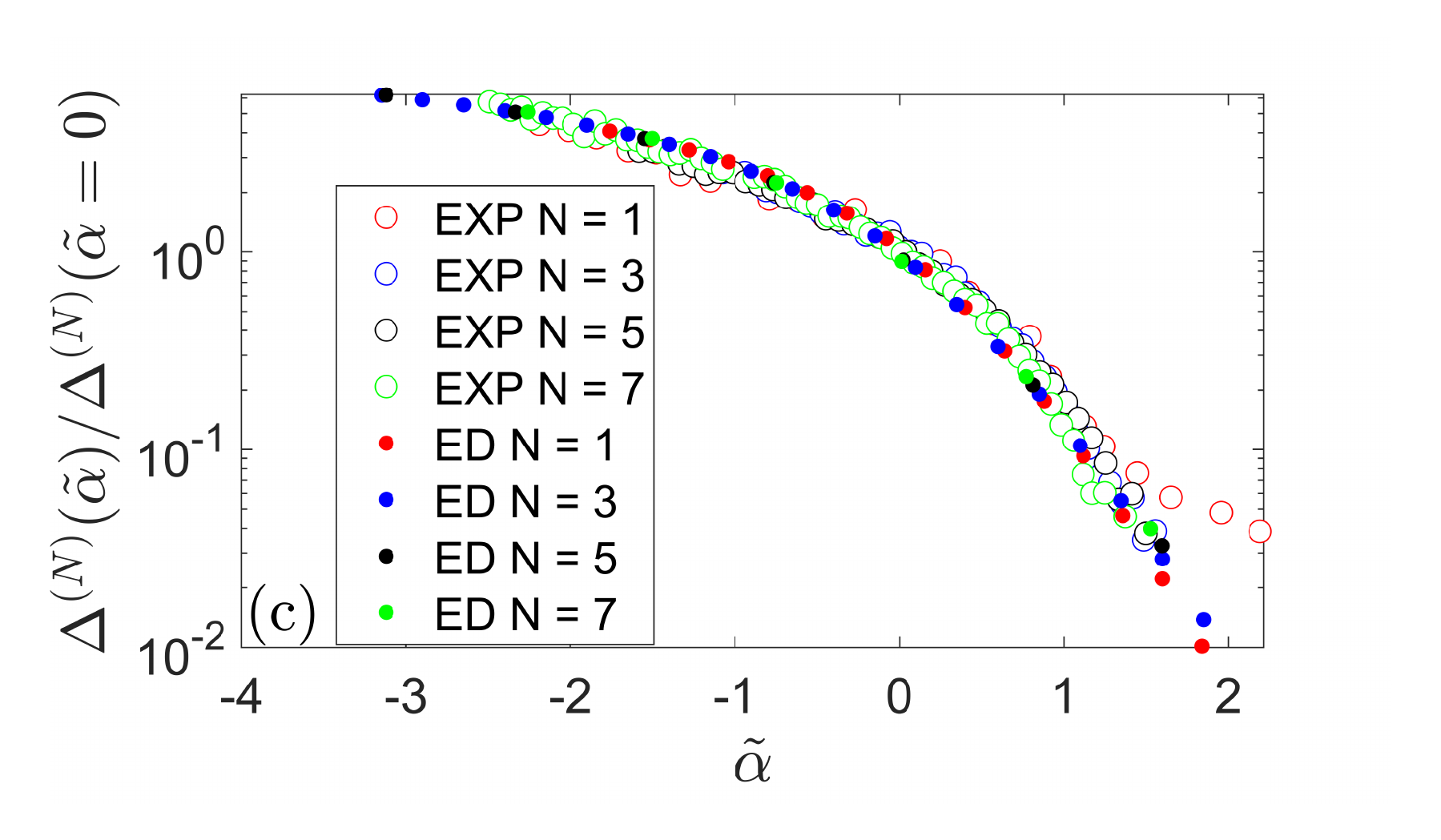}
\caption{(a,b)  Numerically computed tunnel splitting on both linear  and logarithmic scales, compared 
with the rescaled experimental polarization data  (EXP).  Instanton theory captures  $\Delta$  accurately 
in the deep tunneling regime, where $\Delta$ is  suppressed exponentially by increasing barrier height (see  panel (b)).
 Each set of curves corresponds to a different number of electrons, $N$. \zg{The shaded region indicates the regime, 
 where the exchange couplings may already modify the tunneling process and spin can play an important role.}
 In panel (c), the data are scaled together and trace a single,   universal curve. 
 The tunneling regime starts at $\tilde\alpha\approx 0$.
  }   
     \label{fig:EXP_DMRG_ED_IT}
    \end{center}
\end{figure}    

For our computations,  we utilized the Budapest-DMRG code~\cite{Ors,legeza1996accuracy,Ors2016},
which allows us to treat long-range interactions efficiently and to take advantage of the $U(1)$ 
symmetry of the model associated with the conservation of the total charge as well as  the $Z_2$ 
symmetry associated with  parity. In our computations, we use a bond dimension of the order of   2048-4096,  and
an adaptive basis consisting of  $8-16$ 
orbitals/electron, depending on the number of electrons. 
We  computed the ground state energies in the even and odd parity sectors,  $E_{GS}^{(e)}$ and $E_{GS}^{(o)}$, and 
extracted the energy splitting
\begin{gather}
	\Delta  = |E_{GS}^{(e)}-E_{GS}^{(o)}|,\label{eq:gap}
\end{gather} 
identified as twice the tunneling amplitude in the tunneling regime. 
Typical results for $\Delta$ as well as  a comparison with the results of the  other  approaches 
we used are displayed in Fig.~\ref{fig:EXP_DMRG_ED_IT} as function of $\alpha$. Unfortunately, 
in the deep tunneling regime, where     $\Delta $ becomes
exponentially small, $\Delta\lesssim 10^{-3}$, 
we noticed that the convergence of DMRG was influenced by the choice of basis we utilized. Specifically, as we decreased $\alpha$, the states became increasingly localized, which led to convergence challenges for the desired accuracy. Although increasing the bond dimensions improves the calculations, it also demanded greater computational resources. Consequently, as an alternative approach, we employed complementary methods like restricted exact diagonalization or instanton theory to achieve accurate results.
Nevertheless, the range of applicability of DMRG overlaps with that of these methods, and 
enables us to obtain a complete description of the collective tunneling. 

\subsection{Restricted Exact Diagonalization}\label{sec:ED}
As a third,  complementary method, we also used the restricted exact diagonalization (ED),  which can be  utilized to determine the 
 eigenstates of relatively small quantum systems. Here we also  use it to benchmark the other two,  more sophisticated 
 methods.
In this work, we  diagonalize the Hamiltonian~\eqref{eq:Hamiltonian_2} in real space. 


For $N\in \{ 1,3\}$,  diagonalization is performed with a relatively large number of states,  $\sim 100$ for each particle, 
ensuring accurate ground state and a few excited state energies. However, for $N\in \{5,7\}$, the Hilbert space 
becomes too large, and a complete diagonalization  is impossible in practice. 
Nevertheless, a projected version of ED can be used even in these cases, 
where a restricted wave function is used, with electrons treated as distinguishable particles. 
This method reduces the size 
of the Hilbert space by 2-3 orders of magnitude, and can be used to study  up to $N=7$ electrons
 the low-energy spectrum  in the strongly interacting limit, 
where exchange processes are unimportant~\cite{SIMON1984101}.

\section{Results and comparison with experimental data}
\label{sec:comparison}

\subsection{Polarization, polarizability, and tunneling amplitude}

\begin{figure}[t!]
		\begin{center}
			\includegraphics[width=0.49\columnwidth]{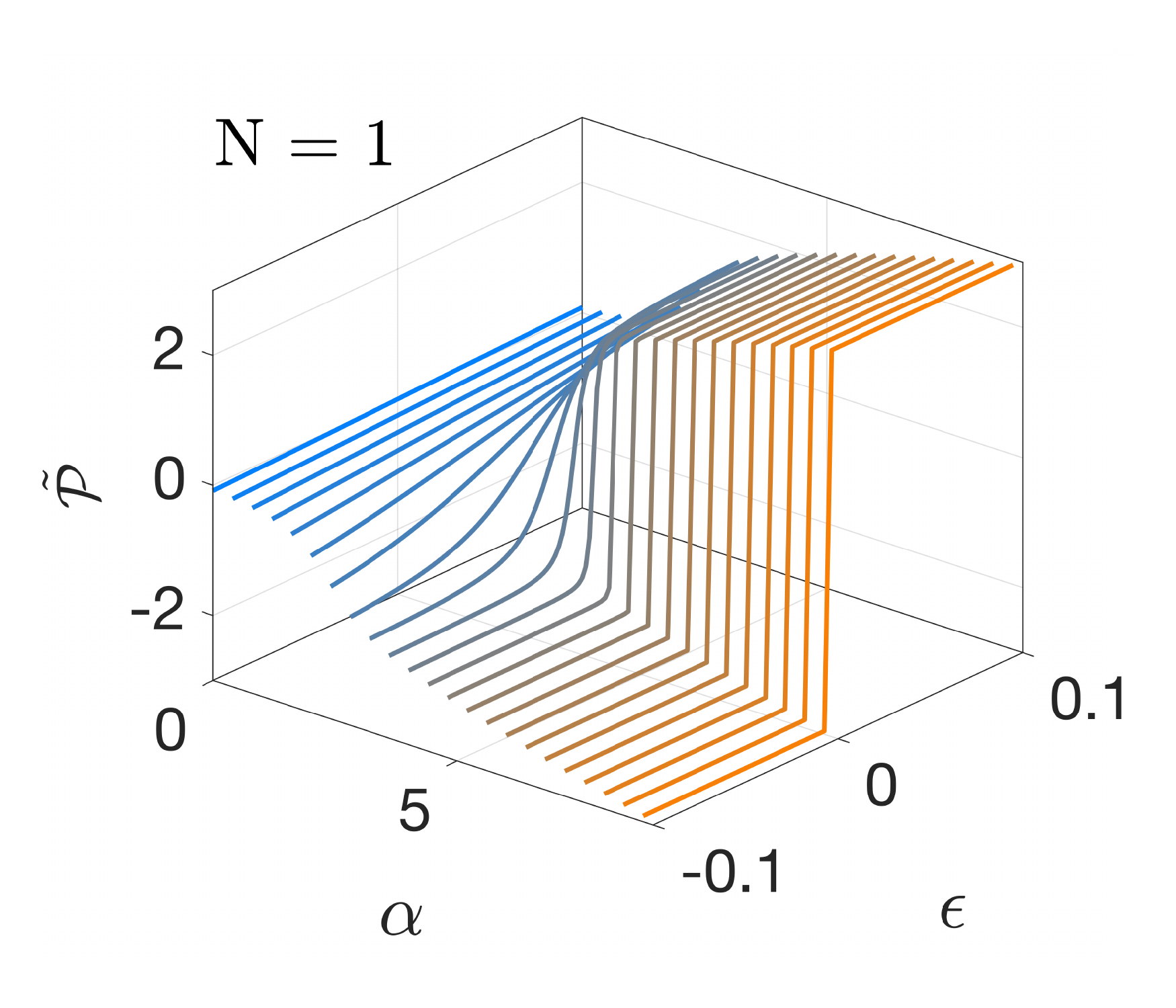}
			\includegraphics[width=0.49\columnwidth]{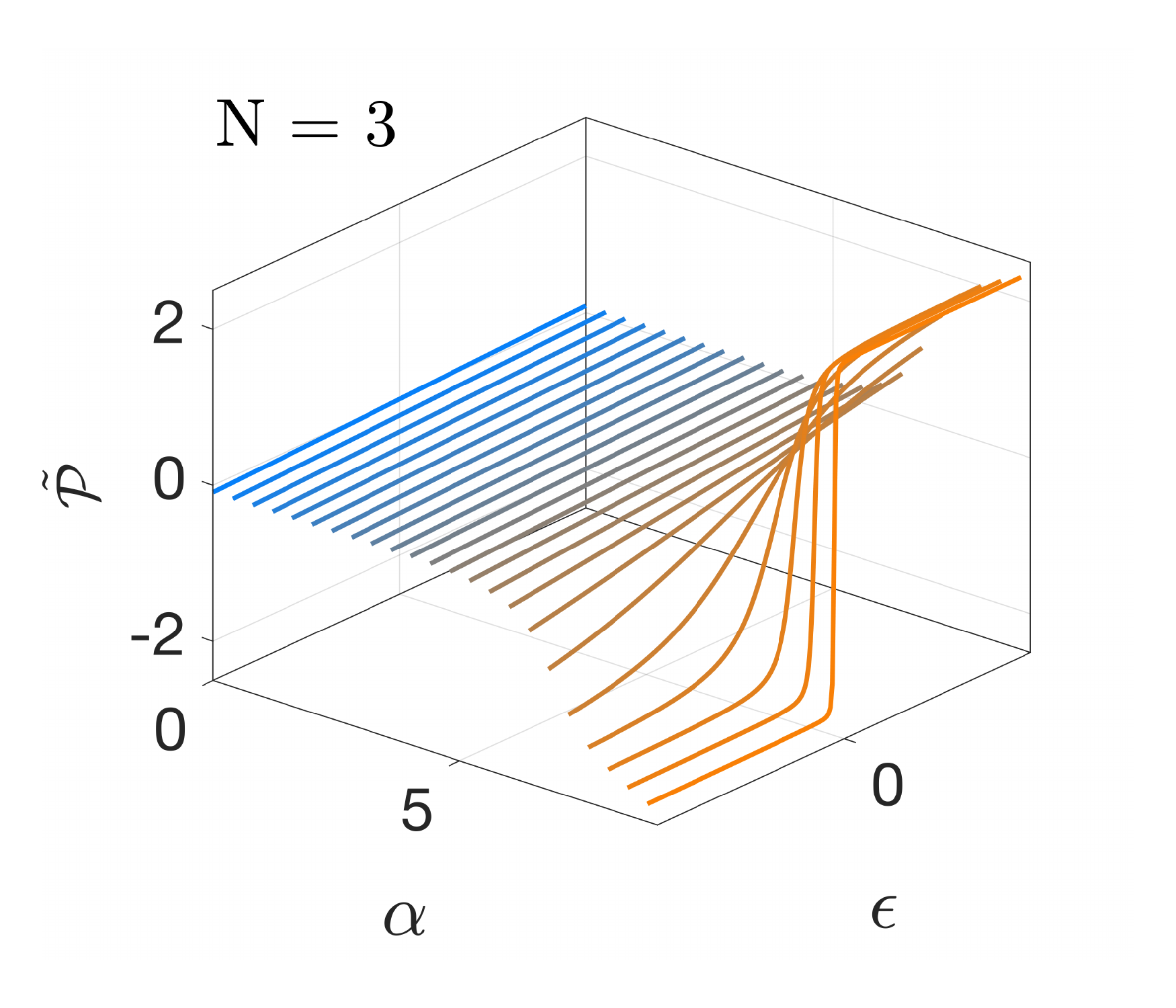}
			\includegraphics[width=0.49\columnwidth]{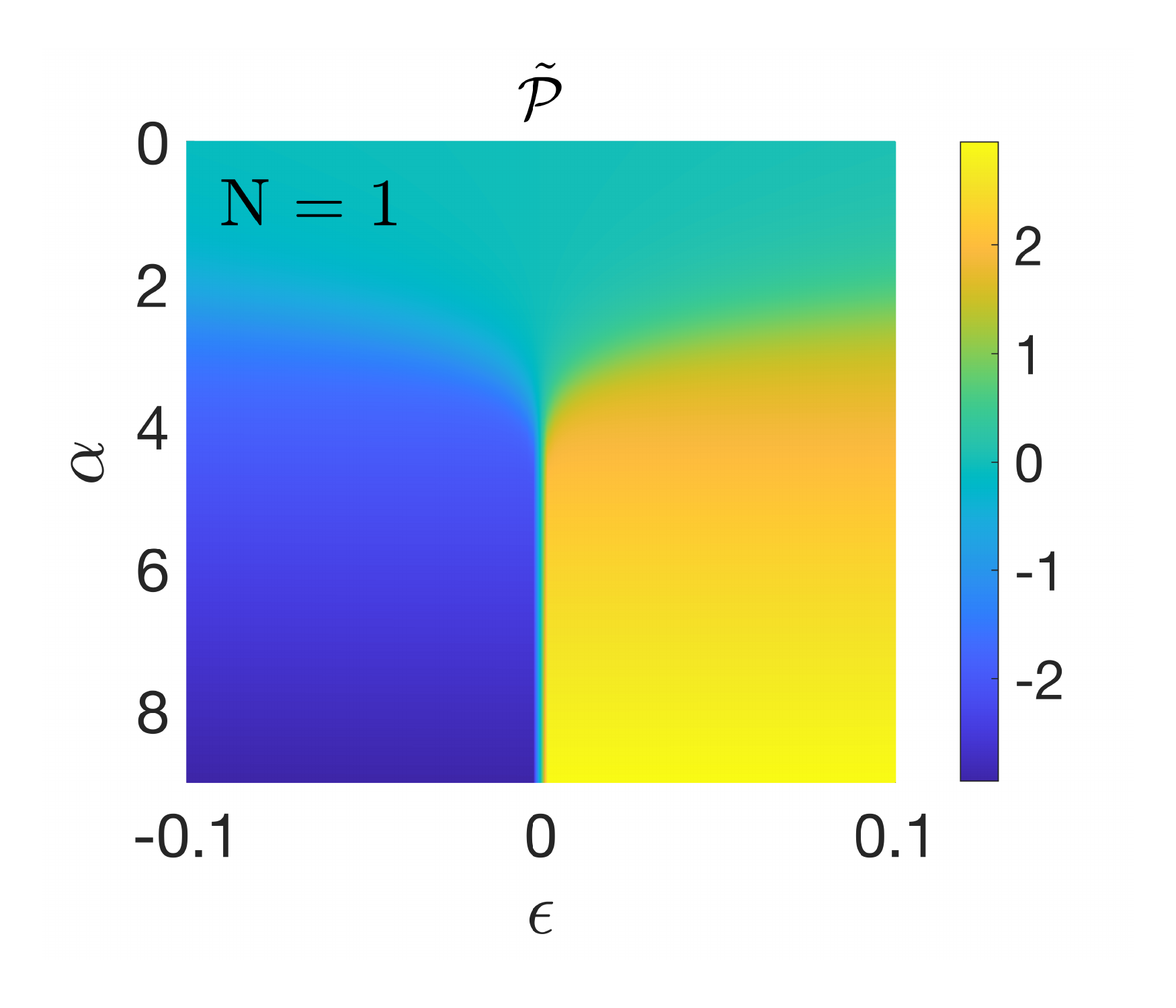}
			\includegraphics[width=0.49\columnwidth]{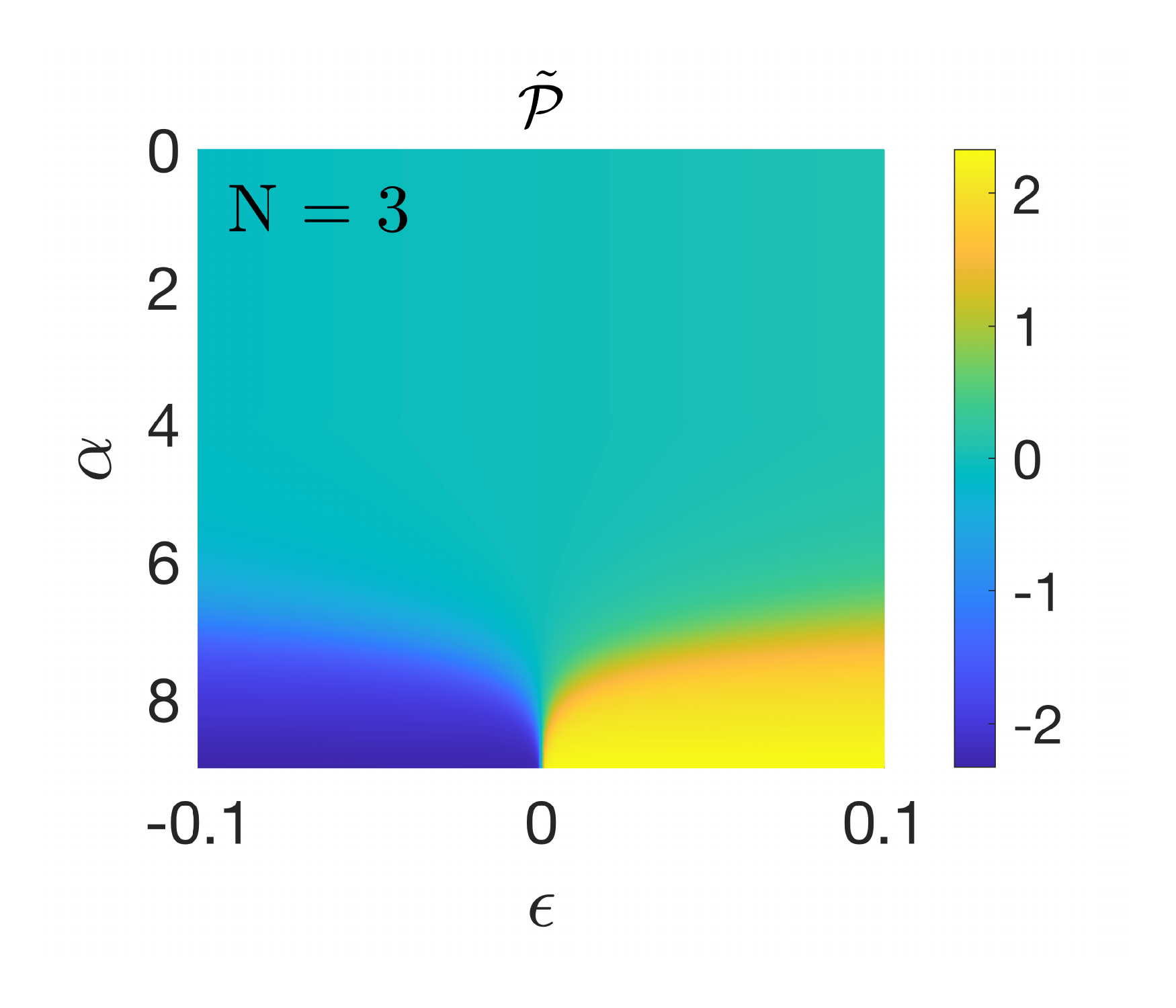}
		 \caption{
		 \label{fig:polarization_comparison}
		 Density plots for the charge polarization $\tilde P(\alpha, \epsilon)$ represented as a function of $\alpha$ and $\epsilon$. The polarization is calculated for $N=1$ and $N=3$ electrons inside the confinement potential.}
		\end{center}
\end{figure}

The careful design and control in the experiments of Ref.~\cite{Shapir.2019} allows one to 
measure the polarization ${P}$ of the electrons on the nanotube as a function of the  applied bias 
($\epsilon$  in Eq.~\eqref{eq:Hamiltonian_2})  as well as  that of the height of the barrier, regulated by a back gate potential $V_K$. 
Such polarization results are displayed in Fig.~\ref{fig:polarization_comparison} along with our theoretical calculations for $N=3$.

In the theoretical computations in Fig.~\ref{fig:polarization_comparison}, we define the dimensionless polarization simply as
\begin{equation}
\tilde{P} (\alpha, \epsilon) = N \langle \chi\rangle = \int\limits {\rm d} \chi  \,\chi \,\rho (\chi, \alpha,\epsilon)   \;,
\label{eq:pol}
\end{equation}
with $ \rho (\chi, \alpha,\epsilon) $ the ground state charge density
\begin{equation}
\rho(\chi,\alpha, \epsilon) = \sum_{i=1}^\n \bra{\Psi}  \delta(\chi - \chi_i)\ket{\Psi},
\end{equation}
and  $\ket{\Psi} = \ket{\Psi(\bchi, \alpha, \epsilon)}$ the ground state wave function 
obtained using ED or DMRG.

As one enters the quantum tunneling regime, the polarization displays a 
kink as a function of the  applied bias. This kink becomes sharper and sharper 
as  the barrier height increases,  clearly demonstrating that 
the broadening of the polarization jump in the experiments is not due to thermal fluctuations, but is 
dominated by \emph{quantum fluctuations} – excepting the very deep tunneling regime, where 
the transition becomes very sharp and its width is set by thermal fluctuations. 

In this quantum tunneling regime,  right at the transition, $\epsilon=0$, 
the polarizability  is inversely proportional to the tunneling amplitude, 
\be
\Pi = \frac{\partial {P}}{\partial \epsilon} \propto \frac 1 {\Delta}\;.
\ee
The precise prefactor here is  hard to determine, since it depends on the precise charge distribution 
before and after the tunneling. Also, although the response at the charge sensor is certainly 
proportional to the polarization, it depends on the capacitive coupling between 
the electrons at various positions and the charge sensor.
Nevertheless, the relation above  enables us to extract the tunneling amplitude as a function 
 of the shape of the barrier, apart from an overall scale. 

For a detailed comparison with the experiments,  we assumed that  there is a linear relation
between the voltage  $V_K$ and  the dimensionless parameter $\alpha$. This leads to the relations
\begin{eqnarray}
	\alpha^{(N)}  &=& \alpha_0^{(N)} +x^{(N)} \; (V_K - V_{0}^{(N)}),\nonumber\\
	\Delta^{(N)} &=&y^{(N)} \; \frac {\Pi^{(N)}(V_{0}^{(N)})} {\Pi^{(N)} (V_K)} .
	\label{eq:rescaling}
\end{eqnarray}
Here the parameters $\alpha_0^{(N)}$ 
and $V_{0}^{(N)}$ mark the threshold of tunneling regime, while $x^{(N)}$ and $y^{(N)}$ rescale the 
axes. We obtain a remarkably accurate fit
to the experiments, as displayed in Fig.~\ref{fig:EXP_DMRG_ED_IT}. Our fitting parameters are 
 enumerated  in Table~\ref{table:parameters}; both $\alpha_0^{(N)}$ and $x^{(N)}$ scale roughly 
 linearly with the threshold, $V_0^{(N)}$, while the overall  polarizability rescaling coefficient
 scales as $y^{(N)}\sim 1/N$.
 Interestingly, the data obtained for various $N$ values also display  a universal scaling when plotted as a 
function of $\tilde \alpha = \alpha-\alpha_0$, as demonstrated in Fig.~\ref{fig:EXP_DMRG_ED_IT}. 
 
%
%
\begin{table}[t!]
	\begin{tabular}{|c|c|c|c|c|}
	\hline
	$N$ & $\alpha_0^{(N)}$   & $x^{(N)} \; [\text{meV}]^{-1}$ & $y^{(N)}\;\text{[meV]}$ & $V_0^{(N)} \;\text{[meV]}$ \\ \hline
	1 & 2.2 & 14.8   & 1    & 109     \\ \hline
	3 & 6.9  & 74.2  & 0.25 & 280     \\ \hline
	5 & 10.4 & 141.9  & 0.15 & 560    \\ \hline
	7 & 13.2  & 170.8  & 0.13 & 850    \\ \hline
	\end{tabular}
	\caption{List of rescaling parameters appearing in Eq.~\eqref{eq:rescaling}.
	}
	\label{table:parameters}
	\end{table}

\begin{figure}[t!]
    \begin{center}
     \includegraphics[width=\columnwidth]{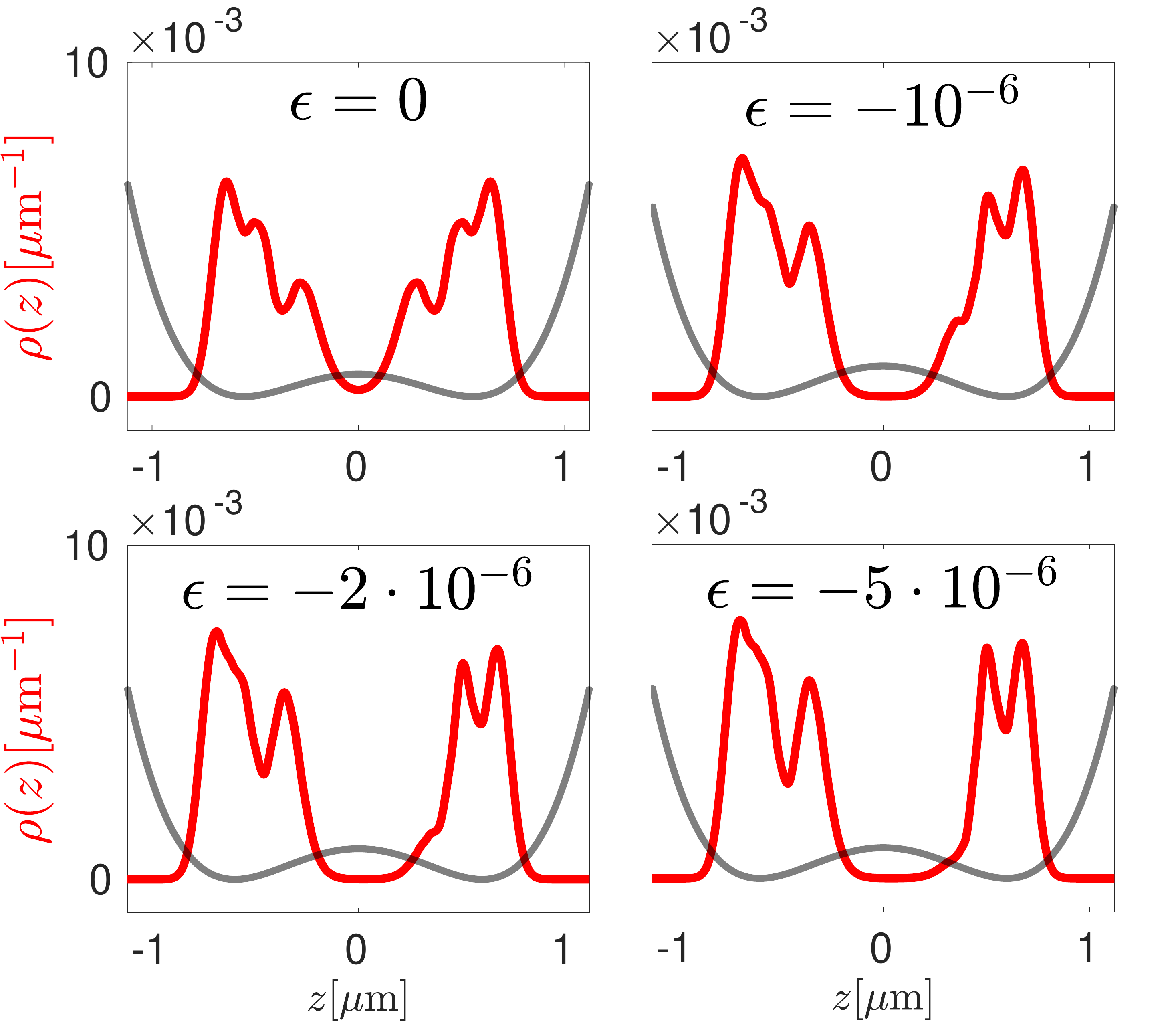}
     \caption{
      \label{fig:charge_distribution_n5} Charge distribution profiles for $N = 5$ particles, with different $\epsilon$ values. At $\epsilon = 0$ the particle in the middle is represented on both sides of the potential barrier, while by increasing the linear detuning, we see that it slowly shifts to one side, changing the positions for the side particles as well.}
	  \label{fig:charge_redistr_n5}
  \end{center}
\end{figure}

\subsection{Charge distribution and polarization}

The experimental set-up of Ref.~\cite{Shapir.2019} also allows measuring charge distributions. 
In particular, the collective motion of the electrons has been demonstrated 
by  measuring the difference of the charge density, $\Delta \rho(z,\alpha, \epsilon)$, 
before and after the tunneling,  and comparing the results with theoretical computations for $N=3$ (see Fig.~4 in 
Ref.~\cite{Shapir.2019}).

\begin{figure}[t!]
\begin{center}
\includegraphics[width=0.49\columnwidth]{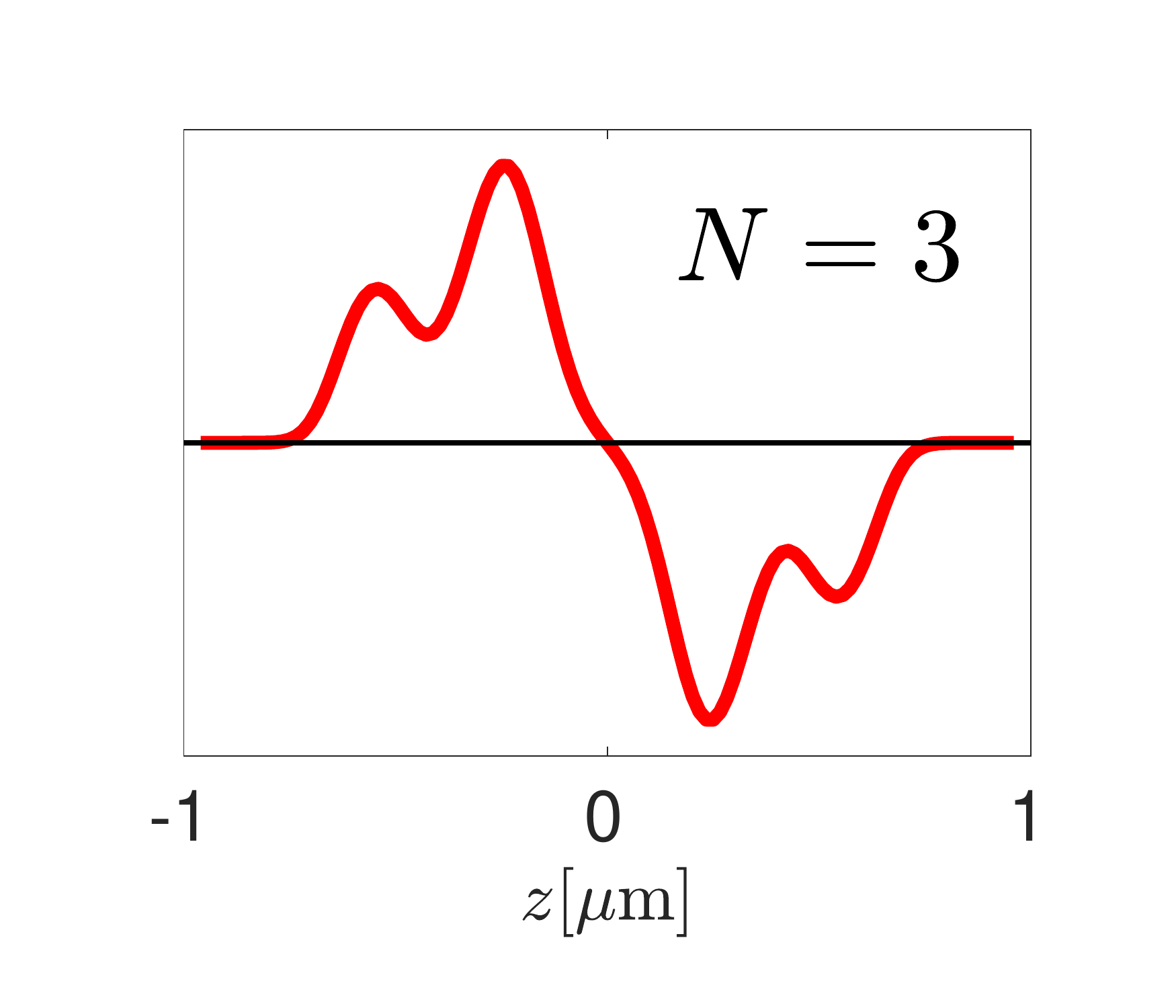}
\includegraphics[width=0.49\columnwidth]{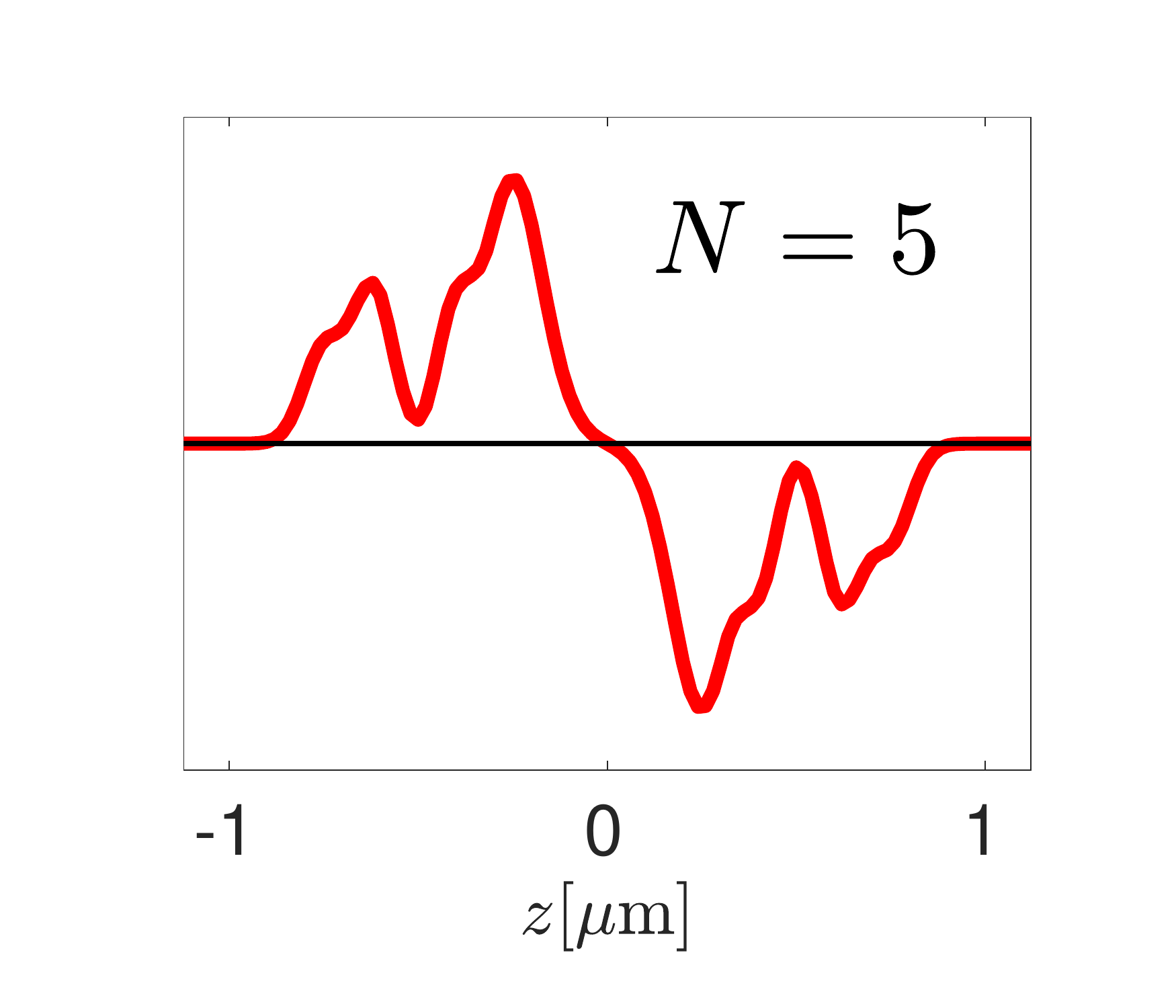}
 \caption{
 \label{fig:charge_conrast}
Smeared  charge density difference, $\Delta \rho(z)$, between the left and right polarized states 
 for $N=3$ and $N=5$ as a function of the bias, $\epsilon$. The peaks emerging on the sides indicates, that during the tunneling all the three particles change their positions. }
\end{center}
\end{figure}

Although experimentally it is not possible to measure non-invasively $ \rho(z)$ at the most interesting point, 
$\epsilon = 0$, we can compute $ \rho(z)$  for any value of $\epsilon$, and study its evolution 
upon changing the bias $\epsilon$. The redistribution of charge as a function of bias, as obtained 
via ED computations is displayed in Fig.~\ref{fig:charge_redistr_n5}  for $N=5$ particles. 
While for $\epsilon \gg \Delta$ two electrons reside on the right and three on the left, 
for $\epsilon\approx 0$ the system delocalizes 
between the "2+3" and "3+2" states, as reflected by the deformation of the density profile. 
While the motion of the central electron is certainly dominant, 
the profile difference, $\Delta \rho(z)$, presented in Fig.~\ref{fig:charge_conrast},
clearly indicates that all charges are displaced in course of the quantum tunneling process.
 
 We present  $\rho(\chi)$  for $N=3$ particles in Fig.~\ref{fig:charge_distribution}
for a set of parameters $\alpha$. The classical ground state  becomes twofold degenerate at  
$\alpha=\alpha^{N=3}_\text{cl}\approx 4.45$. Quantum fluctuations, however, shift this threshold
to $\alpha^{N=3}_\text{0}\approx 6.9$, and tunneling takes place only for  $\alpha \gtrsim  6.3$. 
Fig.~\ref{fig:charge_distribution} also displays the charge polarization, Eq.~\eqref{eq:pol}. 
The main contribution to  the polarization comes from the tunneling of the central electron, 
and the rearrangement of  electrons on the right and the left yields a much smaller contribution. 
At $\epsilon\approx 0$, the central electron is strongly delocalized between the two sides, while the 
lateral electrons gradually shift as the central charge is transferred.

This seems to suggest that lateral electrons are merely spectators of the tunneling event. 
This is, however, not true. As already discussed above, the profile $\Delta \rho$ clearly demonstrates that 
\emph{all electrons} participate in the tunneling process. This is corroborated 
by our instanton computations, which show that lateral electrons participate in 
collective vibrations and thereby  enhance quantum fluctuations,  
which largely facilitate the quantum-tunneling process, as captured by the 
increased prefactor $R_0$ in Eq.~\eqref{eq:instanton}.

		
\section{The role of spin and valley degrees of freedom}\label{sec:spin}
\label{sec:spin}

\begin{figure}[b!]
	\begin{center}
	\includegraphics[width=1.0\columnwidth]{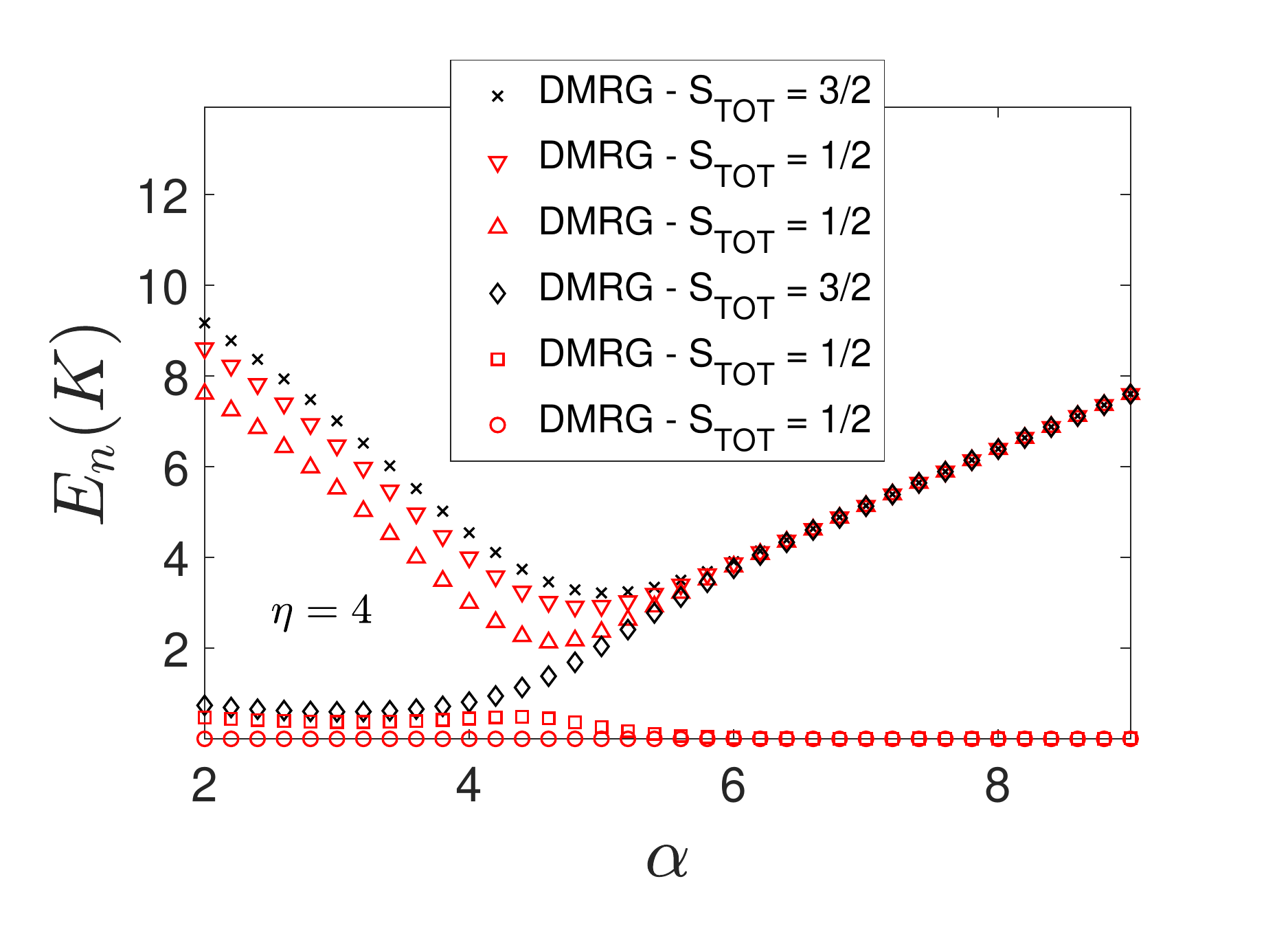}
	\includegraphics[width=1.0\columnwidth]{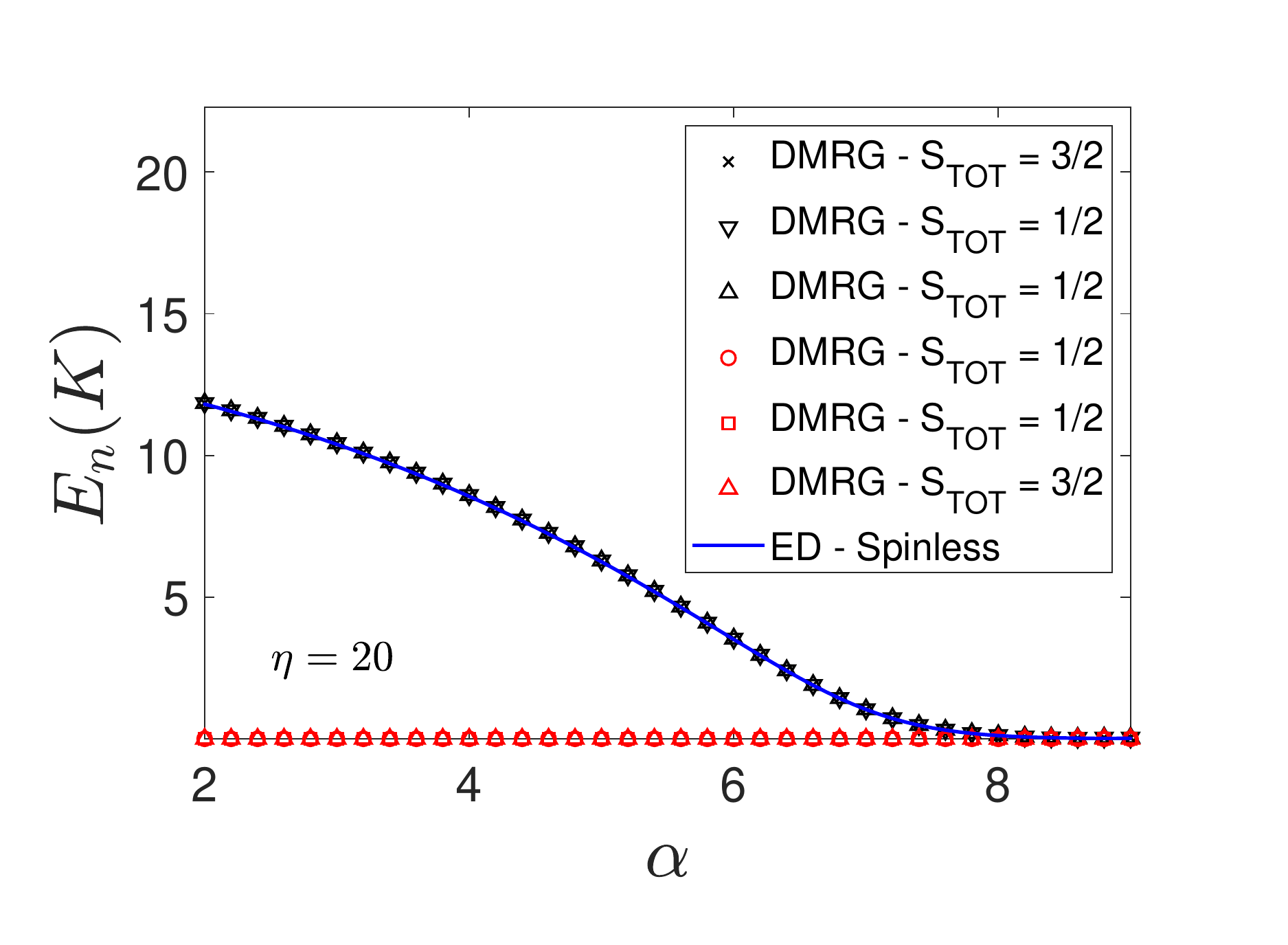}
	 \caption{
	 \label{fig:En_spinhalf}
	 \cpm{The energy spectrum of three electrons with spin confined in a double well potential is depicted, illustrating the influence of spin degrees of freedom under both (a) weak Coulomb interactions with $\eta=4$ and (b) the strong coupling limit with $\eta=20$, as a function of the parameter $\alpha$. } 
	 }
	\end{center}
\end{figure}

\zg{
In this section, we investigate the role of the spin degrees of freedom, and demonstrate that they do not play a 
significant role in the experimentally accessible regime for large interactions, $\eta\approx 20$. 
In addition to spin, electrons in nanotubes also possess a chirality (spin) \cite{KouwenhovenRMP}. 
These play a  role similar to ordinary spins in nanotubes. In the small diameter nanotubes used  in \cite{Shapir.2019}, however, 
spin-orbit coupling is strong~\cite{KouwenhovenRMP}.  Therefore, at the energy scales investigated here, 
 chiral quantum numbers  are already merged with spin degrees of freedom in to a composite $\mathrm{SU}(2)$ degree 
 of freedom, replacing the electrons' ordinary spin \cite{Sarkany.2017}. Spin in the following refers to this 
 composite degree of freedom.}

\zg{
Studying this large interaction range, $\eta\sim 20$
is extremely demanding, and we are not aware of any reliable computation in this regime. 
In this regime, a much larger and adaptive basis \cite{Shapir.2019} is needed  
 to capture the physics at all relevant length scales and reach the 
accuracy requested to determine the size of (exponentially small) level crossings. 
Here we therefore focus only on the case of $N=3$ and 
include only the electrons' composite spin, i.e., we focus on small diameter nanotubes and 
energy scales below a few Kelvins.
}

\cpm{
 The evolution of the ground state energy and that of a few excited states is computed by varying the potential height $\alpha$. 
 The spinful DMRG approach~\cite{legeza1996accuracy} is employed as the method of analysis, utilizing an over-complete 
 harmonic oscillator basis, \zg{centered at around the classical equilibrium positions before and after tunneling \cite{Shapir.2019}}. 
 The calculation employs a total of 24 spinfull orbitals, and a DMRG bond dimension $M=2048$, guaranteeing an accuracy of around 
$\sim 10^{-5}$ in terms of the Schmidt values, corresponding  to an energy precision of around 
\zg{$\sim 10^{-2} \,\rm K$}. }

\cpm{
Fig.~\ref{fig:En_spinhalf} demonstrates  the energy spectrum's dependence on $\alpha$ for intermediate ($\eta=4$, panel (a)),
and strong ($\eta=20$, panel (b)) interactions. 
For intermediate interactions (panel (a)), the spin degrees of freedom  play a significant role:  the ground state is a \zg{doublet}
in agreement with prior results~\cite{Yannouleas.2022}. }
\zg{
The excitation spectrum observed can be interpreted very differently for small and large values of $\alpha$. 
In the limit of small $\alpha\lesssim 5$, the excitation 
spectrum can be understood as a spectrum of a small, $N=3$ Wigner necklace: in this case the electrons form a small spin 
chain with a doublet ground state, and their spin excitations are well described by the effective Hamiltonian, 
$$
H_\mathrm{eff} = J({\bf S}_1 + {\bf S}_3)  {\bf S}_2 \;,
$$
 The ground state and the firts excited states of this Hamiltonian are doublets, separated by the exchange energy $E_1-E_0=J$, while the 
 highest excited state is a spin $S=3/2$ multiplet, located at an energy $E_2-E_0 = 3 J/2$, in good agreement with the 
 spectrum presented in Fig.~\ref{fig:En_spinhalf}, and an exchange interaction, $J\approx 0.5\,\mathrm{K}$. }
\zg{
The excited states at around  $8 \mathrm{K}$ exhibit a strikingly similar spin structure. They can be interpreted 
as a collective vibrational charge excitation, decorated by the spin excitations of the Wigner necklace, 
with a  slightly increased  value of the exchange 
coupling, $J$.}

\zg{The interpretation of the  $\alpha\gtrsim 5$ spectrum for $\eta=4$ is, however, rather different. At large $\alpha$, two electrons are forced to reside on one 
side of the barrier, and the strength of the potential that squeezes them together increases linearly  with $\alpha$. 
As a result, for $\eta =4$ and $\alpha \gtrsim 5$, these two electrons do \emph{not} form a Wigner molecule, but rather, they  occupy 
 the ground state of the confining potential, forming a singlet. These two electrons can tunnel between the r.h.s. and the l.h.s. of the barrier, 
yielding an even spin $S=1/2$  ground state and an almost degenerate odd state of spin $1/2$. The splitting between these states can
 be identified as the quantum tunneling amplitude. }
 
\zg{ The other, higher energy excited states  for $\alpha\gtrsim 5$ and $\eta=4$  can be explained similarly, 
 but in this case one of the electrons resides on the excited state, and forms  a triplet by Hund's rule. This triplet 
 can form a spin $S=1/2$ and a spin $S=3/2$ state with the electron on the other side of the barrier, and form 
 even and odd states, yielding  a group of four excited multiplets, two of $S=3/2$, and two other of $S=1/2$. }

\zg{
In the strong coupling limit, $\eta=20$, the spectrum changes substantially (see   Fig.~\ref{fig:En_spinhalf}(b)). 
In this regime, the exchange couplings are suppressed, and electrons form a Wigner molecule, even when confined to 
one side. As a result, charge degrees  of freedom play the dominant role, and the levels form almost spin degenerate 'bundles'.}

\zg{For $\alpha\lesssim 6$, ground state and the seven almost degenerate low-energy excitations   correspond
to the spin states of an $N=3$ Wigner necklace,  consisting of two almost degenerate doublets and a triplet. 
The first excitation 'bundle' can be interpreted as the first collective charge excitation of the molecule, with the spins 
remaining spectators only.}

\zg{
Upon increasing the barrier $\alpha$, the first  excitation, identified for small $\alpha$ as a collective vibrational charge mode, 
becomes softer and softer, and gradually turns into a tunneling mode between the charge states $(N_L,N_R)=(2,1)$ and $(1,2)$. 
Notice that for $\eta\approx 20$, the spins play the role of silent degeneracy labels even for $\alpha\gtrsim 6$. 

 For comparison, we have also displayed  in panel (b)   the energy levels obtained by solving the Schr\" odinger equation for three spinless, interacting 
particles in real space (solid blue line), demonstrating that the electrons' spin is almost redundant, and the amplitude of tunneling is 
determined simply by  charge excitations. }

\zg{These findings confirm that, in the strongly interacting regime, $\eta\approx 20$, 
it is safe to neglect spin degrees of freedom and consider electrons as spinless particles as long as the 
tunnel splitting associated with the charge degrees of freedom remains larger than the exchange splitting, 
which for $\eta=20$ and $N=3$ is approximately $\Delta_{\rm min}\approx 0.03 \,\rm K$. This threshold 
is just below the smallest experimentally observed tunneling amplitude.} 

\section{Conclusion}\label{sec:conclusions}

\begin{figure*}[t!]
    \begin{center}
     \includegraphics[width=1.6\columnwidth]{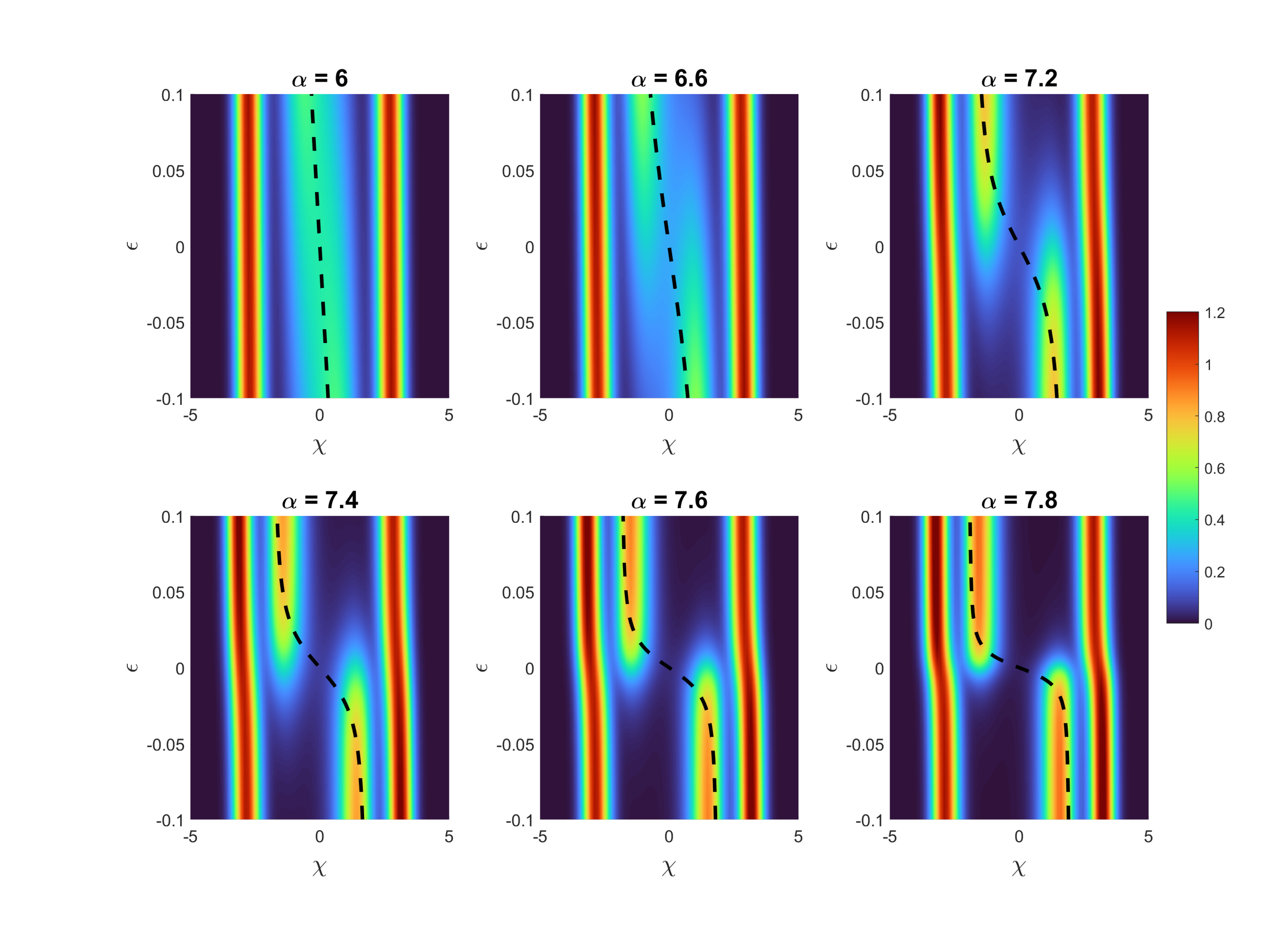}
     \caption{
      \label{fig:charge_distribution}
     Charge distribution (color code) and the polarization (dashed line)  for $N = 3$ particles. The classical critical point is at $\alpha_\text{cl}^{N=3} \approx 4.45$, but tunneling occurs only above $\alpha^{N=3}_0 \approx 6.9$. Most of the polarization is carried by the central electron. Quantum fluctuations of lateral electrons facilitate the tunneling process.}
  \end{center}
     \end{figure*}
In this work, we combined several theoretical approaches to describe the tunneling of 
a \zg{tiny Wigner crystal}  confined within a suspended carbon nanotube and subject to a 
double-well potential, studied experimentally in Ref.~\cite{Shapir.2019}.  For an odd number of electrons
and for sufficiently high barriers,
 the classical ground state of the \zg{Wigner molecule} becomes  degenerate,  and the \zg{necklace} tunnels between these two states. 

A combination of instanton theory, Density Matrix Renormalization Group 
and  a peculiar Exact Diagonalization method allowed us to describe the low energy spectrum of the \zg{necklace}
as well as its charge distribution,  and determine the amplitude of collective tunneling in this very strongly interacting regime, and compare  with the experimental results~\cite{Shapir.2019}.
The methods above provide us  a consistent picture,    compare well with the experiments, and
 provide a quantitative theory for the experimental data in Ref.~\cite{Shapir.2019}.

Interestingly, the tunneling crossover does not take place at the classical bifurcation point, as naively expected, 
but due to quantum fluctuations it is shifted towards somewhat 
higher barrier values (higher values of $\alpha$ in Eq.~\eqref{eq:Hamiltonian_2}). 
Indeed, our calculations clearly demonstrate both the importance of quantum fluctuations 
and the collective nature of tunneling.

Rather surprisingly, we find that the presence of other 
particles \emph{increases} the tunneling amplitude rather than 
reducing it.  
Indeed, in typical tunneling problems the presence of environment leads to a 
\emph{suppression} of tunneling due to Anderson's orthogonality catastrophe \cite{Anderson1967,Leggett1987}.
The physics behind this latter phenomenon is that the motion of one (test) particle influences the 
wave function of all other particles, too, which therefore act back and suppress the motion of 
aforementioned particle. In our case, however,  collective quantum fluctuations of the electron 
chain seem to play a much more important  role: they facilitate the  motion of the innermost electron, 
which is mostly responsible for the tunneling.  This effect is very similar to 
the one found in the case of infinitely long one-dimensional chains, where quantum fluctuations can strongly 
\emph{suppress} the  strength of a pinning center, even in the limit of very strong interactions, 
$r_s \gg 1$, where  pinning  is a strongly relevant perturbation~\cite{Shklovskii.1992,KaneFisher1992}.

Quite astonishingly, the experimental data as well as  our theoretical curves   exhibit a universal 
scaling collapse.  At a first sight, this seems quite natural: one can identify a single collective coordinate 
within the instanton theory, which moves in an effective double well potential, 
and is responsible for the tunneling of the \zg{tiny} crystal. This would support the emergence of
a universal tunneling curve – apart from some overall scaling factors. However,  the remaining 
degrees of freedom  renormalize the tunneling amplitude for $N\ge 3$ particles
by a renormalization factor that has an intrinsic gate voltage dependence.  Apparently, the latter 
renormalization factor, although without an obvious reason,  does not spoil the 
aforementioned universal scaling within  our computational accuracy.

Finally, let us  \zg{discuss}  the role of 
spin and chiral degrees of freedom \zg{in the strongly interacting regime, studied here}. Electrons or holes in a nanotube 
 possess chirality and spin quantum numbers. \zg{For the nanotubes used in the 
 experiments in Ref.~\cite{Shapir.2019}, the large SO coupling freezes spins and chiral spins
 into a single composite spin below a few Kelvins~\cite{Sarkany.2017}. 
We performed DMRG calculations for $N=3$ partciles, where we incorporate the effect of this composite spin.  
We find that  this composite $\mathrm{SU}(2)$ spin plays an important 
 role for moderate interactions, $\eta\sim 5$, in agreement with the results of Ref.~\cite{Yannouleas.2022}, 
 and there it modifies the structure of avoided level crossings,  i.e., the tunneling process.  
 In the strongly interacting regime studied experimentally in  Ref.~\cite{Shapir.2019}, 
 $\eta\approx 20$,  however, we find that spins behave as '\emph{spectators}', and the tunneling process is 
well-described simply in terms of  charge degrees of freedom and spinless particles. 
Neglecting the spin degrees of  freedom is  therefore well-justified when analyzing  
the experiments in Ref.~\cite{Shapir.2019}.}

\dom{In two-dimensional experiments, a strong magnetic field is often applied to shrink the 
cyclotron orbitals, and aid the formation of a Wigner crystal~\cite{Tsui.1982,Goldman1990,Jang2016,Zhou2019}. In contrast, a one-dimensional systems such as  carbon nanotubes,  the magnetic field 
does not compress the orbitals. Rather, it couples to the chirality and the spin of the 
electrons or holes via the Zeeman coupling, and  splits spin and chirality degeneracies~\cite{Sarkany.2017}. 
Transport experiments through the Wigner crystal in the co-tunneling regime in a magnetic field
could  unveil the spin and chirality structure of the ground and excited states. 
}

\zg{
At very low temperatures or smaller interactions, however,  exchange processes 
may become important, and disregarding the spin sector entirely  is not    quite appropriate~\cite{FieteBalents2004,Fiete2007}. 
This may be further complicated by the presence of spin-orbit coupling, especially in larger diameter 
nanotubes: spin-orbit interaction couples spin 
and chiral degrees of freedom, and leads to the freezing  of the charge carriers' SU(4) spin.  
The description of the residual SU(2) degrees  of freedom~\cite{Sarkany.2017}  and their impact 
on the tunneling process at low temperatures  as well as the role of the  SU(4)$\to$SU(2) cross-over 
is an open and very challenging problem, which requires further investigation. }

\acknowledgments
This research is supported by the National Research, Development and Innovation Office NKFIH through research grants 
Nos. K134983, and  
K132146,  
and within the Quantum Information National Laboratory of Hungary (Grant No. 2022-2.1.1-NL-2022-00004). 
%
M.A.W. has also been supported by the Janos Bolyai Research Scholarship of the
Hungarian Academy of Sciences and by the ÚNKP-22-5-BME-330 New National Excellence Program of the Ministry for Culture and Innovation from the source of the National Research, Development and Innovation Fund
C.P.M acknowledges support by the Ministry of Research, Innovation and Digitization, CNCS/CCCDI–UEFISCDI, 
under projects number PN-III-P4-ID-PCE-2020-0277 and the project for funding the excellence, contract 
No. 29 PFE/30.12.2021. 
O.L. has been supported by
Scalable and Predictive methods for Excitation and Correlated phenomena 
(SPEC), funded as part of the Computational Chemical Sciences Program by 
the U.S. Department of Energy (DOE), Office of Science, Office of Basic 
Energy Sciences, Division of Chemical Sciences, Geosciences, and 
Biosciences at Pacific Northwest National Laboratory.
D.S. acknowledges the professional support of the doctoral student scholarship program of the co-operative doctoral program of the Ministry for Innovation and Technology from the source of the National Research, Development and Innovation fund.
\appendix

\section{Computation of the instanton prefactor.}\label{app:prefactor}
Performing the Gaussian integral in Eq.~\eqref{eq:K_full}, one finds \cite{Milnikov.2001}
\begin{gather}
K(\mathbold{\chi}_0^\prime, \mathbold{\chi}_0) = \e^{-S_E} \sqrt{ \frac{2S_E}{\pi}} \left[ \frac{\det^\prime \left(-\partial_\vartheta^2 + V^{\prime \prime}_0 (\vartheta))   \right)}{\det\left(-\partial_\vartheta^2 + 
\omega^2_{\rm{soft}}   \right)}  \right]^{-1/2} \label{eq:K_sep} \\
\times\left[  \frac{\det^\prime \left(-\partial_\vartheta^2 {\bf{1}} + {\bf{\Omega}}^2 (\vartheta)   \right)}{\det\left(-\partial_\vartheta^2 {\bf{1}}+ {\bf{\Omega}}^2_0   \right)}   \right]^{-1/2}.\nonumber
\end{gather}
Here the first line represents the propagator's classical contribution, whereas the second line 
denotes the contribution arising from quantum fluctuations.

The softest vibrational mode at the base of the classical trajectory  $\bchi_{cl}(\tau)$ has a frequency  $\omega_{\rm{soft}}$, and $\det^\prime$ denotes the functional determinant, computed  by excluding the zero eigenvalue in the energy spectrum of the tunneling.

The  $(\n-1) \times (\n-1)$ matrix   ${\bf{\Omega}}_0$ represents the eigenfrequencies around the equilibrium position, $\bchi_0$. The $(\n-1)\times (\n -1)$ matrix ${\bf{\Omega}}(\vartheta)$ is computed using the vibrational eigenvectors along the instanton trajectory. Technically, to compute the contribution coming from the quantum fluctuations is a delicate issue. We followed the approach introduced in Ref.~\cite{Milnikov.2001} and  introduce the Jacobian fields through $\left(-\partial_\vartheta^2 {\bf{1}} + {\bf{\Omega}}^2 (\vartheta)   \right) \bJ(\vartheta)=0$ which is related with the derivative of the instanton $\bkappa(\vartheta) \propto \dot \bchi(\vartheta) $. Introducing a function ${\bf \Xi} (\vartheta) = \dot \bkappa(\vartheta)\, \bkappa(\vartheta)^{-1}$ that satisfies a differential equation,  ${\bf{\dot{\Xi}}} = \varrho /(1 - \vartheta^2)\left({\bf{\Omega}^2}(\vartheta) - {\bf{\Xi}}^2(\vartheta)\right)$, with the boundary condition that the particles behavior is of a harmonic oscillator, we can express the tunnel splitting in a compact form.

An essential aspect of this calculation is the introduction of a coordinate basis transformation on the $\n$ dimensional trajectory. The new basis consists of one parallel and $\n-1$ perpendicular unit vectors with respect to the trajectory, as opposed to $\n$ coordinates that describe the independent particles. It is found that the trajectory's direction is parallel to the eigenvector of the softest vibrational eigenmode.

In this description the trajectory and subsequent calculations can be simplified to an arc-length parametrized effectively one-dimensional description. This takes place in the effective potential, that is created by the collective motion of particles as in Fig. \ref{fig:traj}. This enables us to calculate the quantity $P[{\mathbold{\chi}}_{cl}(s)] $ as a one-dimensional equation. This renormalization constant depends on the momentum-like quantity $p_0 = \sqrt{2 [v_{\n}^{\rm{max}} - v_{\n}^{\rm{min}}]}$
\begin{equation}
P[{\mathbold{\chi}}_{cl}(s)] = \exp\left\lbrace \int\limits_{-1}^0 {\rm{d}}\vartheta\, \frac{\varrho}{(1 - \vartheta^2)} \left( \omega_{\rm{soft}} - \frac{\partial p_0(s)}{\partial s} \right) \right\rbrace.
\end{equation}

 This arc-length parametrized picture makes it possible to express $\mathbold{\Omega}$ by the curvature of the trajectory parametrized either by imaginary time or arc-length, although the two descriptions yield the same results. Solving numerically the set of differential equations ${\bf{\dot{\Xi}}} = \varrho /(1 - \vartheta^2)\left({\bf{\Omega}^2}(\vartheta) - {\bf{\Xi}}^2(\vartheta)\right)$ with the appropriate boundary conditions that states, that indeed for times close to $\vartheta = \pm 1$ the particles behave like a collective harmonic oscillator.

The renormalization factor $R_0(\alpha, \n)$ then can be expressed as
\begin{gather}
	R_0(\alpha, \n) = \sqrt{\frac{\det {\bf{\Omega}}_0}{\det {\bf{\Xi}}(0)}}\times\nonumber \\ \exp \left\lbrace \int\limits_{-1}^0 {\rm{d}}\vartheta \, \frac{\varrho}{(1 - \vartheta^2)}\,{\rm{Tr}}( {\bf{\Omega}}_0 -  {\bf{\Xi}}(\vartheta))\right\rbrace.
\end{gather}

\bibliography{references}

\end{document}